\pdfoutput=1
\documentclass[11pt]{article}
\usepackage[T1]{fontenc}
\usepackage{lmodern}
\usepackage[margin=1.15in]{geometry}
\usepackage{amsmath,amssymb,amsthm,bm}
\usepackage{booktabs,graphicx,setspace,array}
\usepackage[round]{natbib}
\usepackage{xcolor}
\usepackage{mdframed}
\usepackage{listings}
\usepackage[hidelinks]{hyperref}

\newtheorem{theorem}{Theorem}
\newtheorem{proposition}{Proposition}
\newtheorem{lemma}{Lemma}
\newtheorem{corollary}{Corollary}
\newtheorem{remark}{Remark}
\newtheorem{assumption}{Assumption}
\newtheorem{definition}{Definition}

\newcommand{\E}{\mathbb{E}}
\newcommand{\R}{\mathbb{R}}
\newcommand{\tr}{\operatorname{tr}}
\newcommand{\antisym}{\operatorname{asym}}
\newcommand{\KL}{D_{\mathrm{KL}}}

\newcounter{algo}
\newenvironment{algo}[1][]{%
  \par\medskip\refstepcounter{algo}%
  \begin{mdframed}[linewidth=0.5pt,linecolor=black!45,backgroundcolor=black!3,
    innertopmargin=7pt,innerbottommargin=7pt,innerleftmargin=9pt,
    innerrightmargin=9pt]%
  \noindent\textbf{Algorithm \thealgo.}\ \textit{#1}\par\smallskip\small}%
  {\end{mdframed}\par\medskip}

\newcommand{\labTeA}{0.015}
\newcommand{\labDyA}{10.0}
\newcommand{\labGrangerA}{1.00}
\newcommand{\labRevA}{0.03}
\newcommand{\labFeatA}{0.04}
\newcommand{\labQfB}{0.159}
\newcommand{\labSigB}{0.185}

\newcommand{\labDyB}{14.5}

\newcommand{\labRevB}{1.00}

\newcommand{\labQfC}{1.732}
\newcommand{\labSigC}{0.602}

\newcommand{\labGrangerD}{0.13}
\newcommand{\labRevD}{0.04}
\newcommand{\labFeatD}{1.00}
\newcommand{\labCorrGapOrder}{-16}
\newcommand{\labSigAOrder}{-31}
\newcommand{\labMstIdentical}{identical}

\newcommand{\labT}{800}
\newcommand{\labM}{200}
\newcommand{\labR}{499}
\newcommand{\labNsub}{3}
\newcommand{\labCnl}{0.6}
\newcommand{\cexTeGap}{9.18}
\newcommand{\cexEigOne}{0.194}
\newcommand{\cexEigTwo}{0.509}
\newcommand{\cexTeOne}{1.83}
\newcommand{\cexTeTwo}{0.92}
\newcommand{\gridPluginBig}{0.845}
\newcommand{\gridCfBig}{-0.009}
\newcommand{\gridRatioBig}{0.845}
\newcommand{\gridPluginSmall}{0.025}

\newcommand{\gridM}{200}
\newcommand{\gridPluginWorst}{4.41}
\newcommand{\gridCfWorst}{0.28}
\newcommand{\gridRatioMin}{0.76}
\newcommand{\gridRatioMax}{1.31}
\newcommand{\covTrue}{1.324}
\newcommand{\covCf}{1.351}
\newcommand{\covCfBias}{+0.027}
\newcommand{\covCfRelBias}{2.1}
\newcommand{\covPlugin}{2.274}
\newcommand{\covPluginBias}{+0.950}
\newcommand{\covInflate}{1.68}
\newcommand{\covJk}{0.937}
\newcommand{\covM}{300}
\newcommand{\powZero}{0.02}
\newcommand{\powMid}{0.53}
\newcommand{\powMidSig}{0.061}
\newcommand{\powTop}{1.00}
\newcommand{\powTopSig}{0.242}
\newcommand{\powM}{150}
\newcommand{\szLinear}{0.05}
\newcommand{\szFeatAll}{0.05}
\newcommand{\szFeatCross}{0.04}
\newcommand{\szM}{200}
\newcommand{\aliasPhiGapOrder}{-14}
\newcommand{\aliasOmegaOne}{0.8}
\newcommand{\aliasOmegaTwo}{7.08}
\newcommand{\aliasQOne}{-1.00}
\newcommand{\aliasQTwo}{-8.85}
\newcommand{\aliasQDelta}{0.6011}
\newcommand{\aliasRecoverOrder}{-16}
\newcommand{\smallIntSlope}{0.93}
\newcommand{\aggLeakTwo}{0.012}
\newcommand{\aggLeakFive}{0.043}
\newcommand{\aggLeakTen}{0.060}

\newcommand{\tsPeakQ}{0.150}
\newcommand{\tsFastKappa}{16}
\newcommand{\tsFastQ}{0.0003}

\newcommand{\tsSlowQ}{0.074}
\newcommand{\tsCorrDevOrder}{-16}
\newcommand{\aggFastTwo}{0.63}
\newcommand{\aggFastFive}{0.106}
\newcommand{\aggFastTen}{0.014}
\newcommand{\aggSlowTwo}{1.19}
\newcommand{\aggSlowFive}{0.817}
\newcommand{\aggSlowTen}{0.280}
\newcommand{\aggCheckT}{2.72}
\newcommand{\aggCheckK}{12}
\newcommand{\weT}{1200}
\newcommand{\weSigTrue}{0.1854}
\newcommand{\wePlugin}{0.2248}
\newcommand{\weCf}{0.2207}
\newcommand{\weJkSe}{0.0209}
\newcommand{\weJkLo}{0.1797}
\newcommand{\weJkHi}{0.2616}
\newcommand{\weP}{0.008}
\newcommand{\weR}{1999}

\newcommand{\weCtwelve}{0.3784}
\newcommand{\weCtwentyone}{0.5577}
\newcommand{\weQtwelve}{-0.0897}
\newcommand{\weBiasGap}{0.0042}
\newcommand{\weBiasPred}{0.0050}
\newcommand{\weCfDevSe}{1.7}
\newcommand{\weJkSeHalf}{0.033}
\newcommand{\weJkSeDouble}{0.030}
\newcommand{\weNetEast}{-0.071}
\newcommand{\weNetTrueEast}{-0.062}

\newcommand{\appN}{26}

\newcommand{\appPAdjCrOne}{0.008}
\newcommand{\appSigCrTwo}{4.36}

\newcommand{\appPAdjCrTwo}{0.008}

\newcommand{\appPAdjCaOne}{0.008}

\newcommand{\appPAdjCaTwo}{0.023}

\newcommand{\appPCaThree}{0.004}
\newcommand{\appPAdjCaThree}{0.050}

\newcommand{\appPCaFour}{0.006}
\newcommand{\appPluginCrTwo}{10.35}
\newcommand{\appPluginRatioCrTwo}{2.4}
\newcommand{\appNPairs}{325}
\newcommand{\appSigPerPairCrTwo}{13.430}
\newcommand{\attObs}{0.49}
\newcommand{\attPred}{0.17}
\newcommand{\appTeCrisis}{0.0466}
\newcommand{\appTeCalm}{0.0047}
\newcommand{\appTCrisis}{3.38}
\newcommand{\appTCalm}{1.99}
\newcommand{\appPShiftCrisis}{0.0005}
\newcommand{\appPShiftCalm}{0.0175}
\newcommand{\appRatio}{9.9}
\newcommand{\appOfrLo}{3.15}
\newcommand{\appOfrHi}{3.87}
\newcommand{\appFfbs}{100}

\newcommand{\appTailBeta}{0.30}
\newcommand{\appTailT}{3.12}
\newcommand{\appTailBetaCalm}{0.06}
\newcommand{\appSettleTe}{0.263}
\newcommand{\appSettleT}{12.4}
\newcommand{\appSettleInfoT}{1.12}
\newcommand{\appPlcMin}{0.00}
\newcommand{\appPlcMax}{0.06}
\newcommand{\appPlcM}{100}
\newcommand{\adqTrue}{10.35}
\newcommand{\adqCf}{10.39}
\newcommand{\adqPlugin}{11.79}
\newcommand{\adqQCorr}{0.98}
\newcommand{\adqCover}{0.950}
\newcommand{\adqNullPlugin}{0.85}
\newcommand{\adqNullCf}{0.01}
\newcommand{\invMaxRelOrder}{-10}
\newcommand{\invDraws}{200}
\newcommand{\exT}{800}
\newcommand{\exQfrob}{0.1875}
\newcommand{\exPlugin}{0.2242}
\newcommand{\exCf}{0.2169}
\newcommand{\exP}{0.006}
\newcommand{\exNetOne}{-0.026}
\newcommand{\exNetTwo}{+0.027}
\newcommand{\exNetThree}{-0.001}
\newcommand{\exGapOrder}{-13}
\newcommand{\figPRev}{0.54}
\newcommand{\figPCirc}{0.001}

\title{Measuring the Arrow of Time:\\
Identification, Estimation, and Inference for\\
Directional Structure in Multivariate Time Series}

\author{Avishek Bhandari\thanks{avishekb@iitbbs.ac.in.
School of Humanities, Social Sciences and Management, Indian Institute of
Technology Bhubaneswar, India.}}
\date{August 2026}

\begin{document}
\maketitle
\begin{abstract}
\noindent
Many questions across the sciences take the same form: several coupled series
are observed together, and the analyst wants to know not merely that they move
together but which one moves first, and how strongly. This paper sets out a
complete method for answering that question, built on a single organising idea:
the direction of a coupled system is exactly the part of its behaviour that
changes when the record is played backwards. Tools built on contemporaneous
covariance alone, correlation matrices, distance measures, spanning trees,
undirected network centralities, principal components, carry no information
about direction: a reversible system and a circulating one can share identical
covariance at every sampling of the same point-in-time record, and only
temporal aggregation lets a small, computable trace through. What direction is,
formally, is a circulation matrix carried by the lagged covariance. Its
vanishing is exactly statistical time reversibility for linear systems, feature
maps carry the same characterisation to nonlinear ones, and under the Gaussian
benchmark its magnitude is an entropy-production functional of the identified
circulation, the quadratic component of the divergence per unit time between
the forward record and the reversed one. Around this estimand we build an estimator whose
first-order bias is removed by cross-fitting, standard errors that reuse no
observation within a replicate, and a randomisation test that draws its
reference distribution from the null hypothesis's own symmetry, exact under the
stated block structure, with a familywise correction that costs nothing and a
nonlinear extension that leaves the reference distribution intact. A sampling
theory says when the arrow is measurable at all: too fast a system, or too
coarse a record, and it disappears. Because a measured arrow is not yet an
economic claim, a design layer separates transmission from the mere ordering of
clocks. A laboratory of four systems with known answers compares the method
with correlation networks, Granger causality, transfer entropy, and
connectedness indices, and reports the failures of each when read as a measure
of direction, including our own: in a reversible system, predictive and
variance-share measures still report nonzero directional readings, correct for
the questions they answer but routinely misread as an arrow, while in a system
with purely nonlinear transmission our linear reading is blind and the
feature-map extension recovers it. The paper is self-contained, gives complete
algorithms and worked examples in two languages, and is intended as the base
reference for a series of applications.


\textbf{MSC classes:} 62M10, 62G09, 60J60.
\end{abstract}
\newpage

\onehalfspacing
\section{Introduction}\label{sec:intro}

Across the empirical sciences the same question keeps being asked in different
vocabularies. An epidemiologist watching case counts in neighbouring regions
asks which region seeded which. A neuroscientist watching two brain areas asks
which one drives the other. A climate scientist watching temperature and carbon
records asks which leads. An economist watching several national stock markets
asks whether stress in one was transmitted to another, or whether both simply
responded to the same news. In every case the analyst has several series
recorded side by side, sees that they move together, and wants to know
something stronger than togetherness: an ordering.

The standard empirical answer is to compute a matrix of correlations and to
build something on top of it. Correlation matrices become distance matrices,
distance matrices become networks and spanning trees, networks yield
centralities, and centralities are read as influence. This family of methods is
enormously useful for describing which series belong together. This paper
begins with a result showing that it cannot answer the question of direction at
all, not because of noise or sample size, but as a matter of arithmetic: two
systems with opposite directional structure can generate exactly the same
covariance matrix however often or seldom the same point-in-time record is
sampled, forever. Everything computed from contemporaneous co-movement is
therefore constant across systems that differ in direction. The information
simply is not there. (One qualification is worth flagging at the outset:
\emph{aggregating} the record, summing daily returns into weekly ones rather
than sampling the same series more coarsely, leaks a small and exactly
computable trace of the direction into the covariance; Section~\ref{sec:sampling}
gives the formula and its size.)

Our organising idea is old and, we think, clarifying. Suppose you record a
system for a long time and then play the recording backwards. If the reversed
recording is statistically indistinguishable from the original, then the system
has no direction: any claim that $A$ leads $B$ would be equally supported by
the claim that $B$ leads $A$, since the reversed record is just as likely as the
original one. If the reversed recording is distinguishable, the system has a
direction, and the strength of the direction is the degree to which the two
recordings can be told apart. This is the physicists' notion of an arrow of
time, and it is a complete answer to what direction can mean for a system in a
statistical steady state.

Turning that idea into a working method requires four steps, and the four
steps organise the paper.

The first step is to define the directional parameter precisely. For a linear
system in a steady state, the definition is sharp. The dynamics split into two parts that live
in orthogonal directions: a part that pulls the system back towards its
long-run average, which is completely determined by the covariance matrix, and
a part that circulates the system around the contours of its long-run
distribution, which the covariance matrix does not constrain at all. That
second part is a matrix $Q$, antisymmetric by construction, which we call the
circulation. It is the whole of what direction means: $Q$ vanishes exactly when
the system is reversible; it is recovered from the covariance of the series
with its own recent past; and the natural measure of its size is the rate at
which the forward record diverges from the reversed one, which physicists call
the entropy production rate, of which the index we report is the quadratic
component. Section~\ref{sec:framework} states this and
Appendix~\ref{app:proofs} proves it.

The second step is estimation, and its central difficulty is a bias that
flatters the analyst. The parameter $Q$ itself
is a difference of two sample moments and is unbiased. Its natural magnitude is
a quadratic function of $Q$, and quadratic functions of noisy estimates
inherit the noise as a systematic upward bias: with $n$ series and $T$
observations the bias grows like $n^2/T$, and in a study with twenty-six series
and eight hundred observations that bias is not a technical detail but the
entire measured quantity. Section~\ref{sec:estimation} gives the estimator that
removes it, by computing the two factors of the quadratic from disjoint stretches
of time, together with standard errors that never reuse an observation. In a
system whose true magnitude is zero, the naive estimator here returns
$\gridPluginBig$ and the corrected one returns $\gridCfBig$.

The third step is inference that assumes nothing about the distribution of
the data. The null hypothesis of no direction is unusually generous: it says the record
and its reverse have the same law. That is not a parameter restriction to be
tested with an asymptotic approximation, it is a symmetry, and a symmetry
supplies its own reference distribution. Reversing stretches of the observed
record produces new records that are exactly as likely as the original one if
the null is true, so the observed statistic can be compared with its own
reversals. Section~\ref{sec:inference} develops this, along with the two
practical traps: the reversal must be applied to the raw record and not to
anything filtered from it, and a family of such tests can be corrected for
multiplicity at no additional cost because the same reversals serve every
member of the family.

The fourth step is to establish when the arrow is measurable at all. A record
sampled once a day cannot
see a rotation that completes in an hour: the observations are the same as
those of a much slower rotation, a phenomenon known as aliasing.
Section~\ref{sec:sampling} makes this quantitative. Two systems whose rotation
frequencies differ by a full turn per sampling interval leave footprints that
agree to $10^{\aliasPhiGapOrder}$, so no amount of data separates them, and as a
system is made faster at a fixed observation interval its measured arrow
collapses towards zero even though the underlying circulation grows. There is,
in effect, a window of time scales in which direction is visible, and the
analyst should know where the window is before running anything.

Two further sections address matters that a purely statistical treatment tends
to leave out. Section~\ref{sec:design} is about the gap between a measured
arrow and an economic claim. Markets in different time zones trade at different
hours; a shock arriving during the London afternoon reaches Tokyo's recorded
close a calendar day later purely because of the rotation of the earth. That
mechanism produces a genuine arrow of time with no transmission of any kind. No
statistical procedure can distinguish it from transmission, because both are
real properties of the sampled law; only a research design can, and we set out
the design we use, which splits the receiving day at its opening auction.
Section~\ref{sec:comparison} compares the method with the tools it is meant to
complement, on four systems whose answers are known by construction. The
results are not uniformly flattering to anyone. In a system that is exactly
reversible, so that no directional claim of any kind is warranted, a Granger
causality test rejects in $\labGrangerA$ of samples and a connectedness index
reports a net directional flow of $\labDyA$ per cent; both are answering their
own questions correctly, and both are routinely read as answers to ours. In a
system where transmission is purely nonlinear, our own linear reading is
exactly zero and misses it entirely, which is why the paper also develops a
feature-map extension that detects it in $\labFeatD$ of samples while leaving
the exact reference distribution untouched.

The individual ingredients have histories, and the debts deserve to be stated
plainly. The decomposition of a linear diffusion into a gradient part and a
circulating part is standard in statistical physics
\citep{kwon2005structure,godreche2019characterising}, the equivalence of
reversibility with detailed balance is classical probability
\citep{kelly1979reversibility,jiang2004mathematical}, and entropy production is
a mature subject \citep{seifert2012stochastic,roldan2010estimating}. Measuring
broken detailed balance from observed trajectories is an active experimental
programme in biophysics
\citep{battle2016broken,gnesotto2018broken,martinez2019inferring,seif2021machine},
where the finite-sample bias of plug-in current estimators is a recognised
problem \citep{frishman2020learning}, and the same estimand has recently been
carried into neuroscience \citep{lynn2021broken}. Reversibility of Gaussian
processes is classical \citep{weiss1975time}, and testing reversibility has a
literature of its own in time series
\citep{lawrance1991directionality,ramsey1995time,chen2000testing,
darolles2004tests,racine2007versatile,beare2014time}. The identification of a
continuous-time system from discretely sampled data, and the aliasing
obstruction, are due to \citet{phillips1973problem} and
\citet{hansen1983dimensionality}. The decomposition of linear dependence into
directional feedbacks begins with \citet{geweke1982measurement}.
Randomisation inference is textbook material \citep{lehmann2005testing}, as is
stepdown multiplicity control \citep{romano2005stepwise}.

What is assembled here, and what we have not found elsewhere, is the
combination: a directional estimand with a proof of what it can and cannot
identify; a magnitude with its finite-sample bias characterised and removed; an
test that is exact under its stated block null, with no asymptotic content; a
multiplicity correction obtained from
the same draws; a nonlinear extension that preserves that exactness; a sampling
theory that tells the analyst when the exercise is feasible; a design layer
that separates mechanism from transmission; a comparison against the standard
toolkit on systems with known answers; and complete software in two languages.
Sections~\ref{sec:framework} to \ref{sec:implementation} are written so that
each of these can be used on its own.

The empirical example that runs through the paper concerns twenty-six financial
markets, and Section~\ref{sec:application} summarises it. Nothing in the method
is specific to finance. We use finance because it supplies the hardest version
of the identification problem we know: dozens of coupled series, a genuine
mechanical arrow from the rotation of the earth, regimes that change, heavy
tails, volatility clustering, and a literature that has been reading direction
off correlation networks for three decades.

The remainder of the paper proceeds as follows. Section~\ref{sec:framework}
sets out the framework: the picture behind the estimand, the identification
results, the magnitude and its interpretation, and the nonlinear extension.
Section~\ref{sec:estimation} develops the estimators and their finite-sample
properties, and Section~\ref{sec:inference} the finite-sample inference.
Section~\ref{sec:sampling} gives the sampling theory,
Section~\ref{sec:design} the research design that separates a measured arrow
from a claim about transmission, and Section~\ref{sec:comparison} the
comparison with the standard toolkit on systems with known answers.
Section~\ref{sec:implementation} describes the two reference implementations
with worked examples, Section~\ref{sec:application} summarises the empirical
application, and Section~\ref{sec:limitations} states the limitations and the
open problems. Section~\ref{sec:conclusion} concludes. Proofs are collected in
Appendix~\ref{app:proofs}; Appendix~\ref{app:estimation} contains the
simulation evidence behind every claim about bias, coverage, size, and power;
Appendix~\ref{app:algorithms} states every algorithm in executable detail;
Appendix~\ref{app:laboratory} constructs the four comparison systems;
Appendix~\ref{app:background} is a short primer on the ingredients we assume,
written for readers from outside econometrics; and
Appendix~\ref{app:notation} collects the notation.

\section{Theoretical Framework}\label{sec:framework}

This section develops the theory. It begins with the picture behind the
estimand, then defines the directional parameter, states what it does and does
not identify, gives it a magnitude with an information-theoretic meaning, and
extends it beyond linear structure. Proofs are in Appendix~\ref{app:proofs}.
Readers unfamiliar with stationary linear systems, the Lyapunov equation, or
Kullback--Leibler divergence will find a short primer in
Appendix~\ref{app:background}.

\subsection{Direction as time irreversibility}\label{sec:picture}

Consider two markets whose returns are recorded once a day. Over a long stretch
of time the pair traces out a cloud of points, elongated along a diagonal
because the two markets are positively correlated. That cloud is the system's
long-run distribution, and it is what a correlation matrix summarises.

\begin{figure}[t]
\centering
\includegraphics[width=0.86\textwidth]{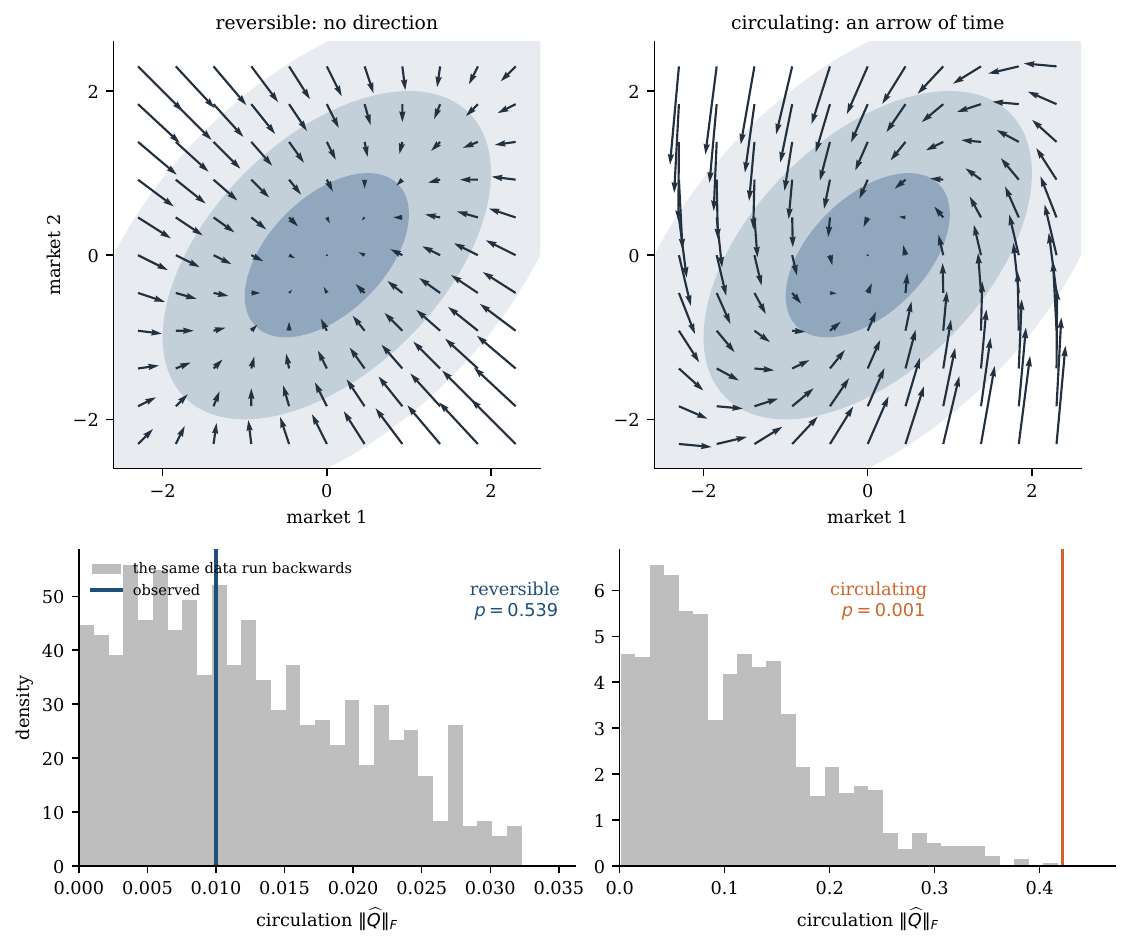}
\caption{Two systems with the same long-run distribution and different
behaviour. Shaded contours are the long-run distribution, arrows are the
average next step from each point. Left: a reversible system, in which the
average step points down the steepest slope of the distribution. Right: the
same distribution with a circulating component added, so that the system turns
as it returns. Bottom row: for a sample of $1{,}500$ observations from each
system, the measured circulation (vertical line) against the distribution of
the same statistic recomputed on the same data with stretches of time reversed
(grey). The reversible system sits inside its own reversals
($p = \figPRev$); the circulating one does not ($p = \figPCirc$).}
\label{fig:direction}
\end{figure}

A correlation matrix does not, however, say where the system standing at a
given point in the cloud tends to move next. Figure~\ref{fig:direction}
displays that movement for two systems
that have exactly the same cloud. In the left panel the typical next step
points straight back towards the centre: the system is pulled home along the
steepest path, and nothing else happens. In the right panel the typical next
step also returns towards the centre, but it leans sideways as well, so the
system drifts around the cloud as it returns. The clouds are identical. The
correlations are identical. The behaviour is not.

The difference is visible only in time. Run the left panel backwards and it
looks the same: a cloud with arrows pointing inward is unchanged by reversing
the direction of travel, because reversing an inward step gives an outward step
which the picture already contained by symmetry. Run the right panel backwards
and the rotation changes sign, which the original picture did not contain. This
is what we mean by an arrow of time, and it is the whole content of the word
direction for a system in a steady state.

The observable consequence is equally simple. Write $C(\Delta)_{ij}$ for the
covariance between market $i$ today and market $j$ one period ago. In the left
panel, $C(\Delta)_{ij} = C(\Delta)_{ji}$: today's market $1$ tells us as much
about yesterday's market $2$ as today's market $2$ tells us about yesterday's
market $1$. In the right panel the two differ, and the difference
\[
Q_{ij} \;=\; \tfrac{1}{2}\bigl(C(\Delta)_{ij} - C(\Delta)_{ji}\bigr)
\]
is a directed flow from $j$ into $i$. That single expression, a difference of
two numbers both of which any spreadsheet can compute, is the estimand of this
paper. Everything else is about doing it properly: what it identifies, how to
measure its size without bias, how to test it without asymptotic
approximations, when the sampling makes it invisible, when a measured flow
means transmission, and where it is blind.

Three points about the picture are worth making now, because they recur.

The picture is symmetric in a way that matters. The left panel is not a
special case with weak dependence. It can have arbitrarily strong correlations
and arbitrarily strong persistence, and one market can be far more persistent
than the other. What it cannot have is a preferred direction of travel. Methods
that read direction from asymmetries in predictability, such as Granger
causality and transfer entropy, do report a direction in the left panel, because
the more persistent market is harder to predict from the other one, and that
difference has nothing to do with direction of travel. We return to this in
Section~\ref{sec:comparison} with a worked instance in which the two transfer
entropies differ by $\cexTeGap \times 10^{-4}$ nats in a system that is exactly
reversible.

The measurement also has a natural null hypothesis. If the system has no
direction, then reversing stretches of the record produces records that are
just as likely as the one observed. So the observed value of any statistic can
be compared with the values it takes on those reversals, and the only
assumption used is the null itself, stated with its block structure in
Section~\ref{sec:inference}. The bottom row of Figure~\ref{fig:direction}
shows exactly this comparison, with the reversible system falling in the middle
of its own reversals and the circulating system falling far outside.

Finally, the size of the arrow has a meaning, not just a scale. A single number
summarising the whole matrix $Q$ can be chosen arbitrarily, but one choice is
canonical: the rate at which an observer accumulates evidence that the record
is being played forwards rather than backwards. That is a divergence per unit
time between two probability laws, measured in nats per period, and for the
linear system of the picture it is a simple quadratic function of $Q$. A value
of zero means the two directions of time are indistinguishable. A value of one
nat per day means a day of data is enough, on average, to shift the odds
between forwards and backwards by a factor of about $e$. This gives magnitudes
an interpretation that does not depend on the units of the data.

\subsection{The directional parameter}

Let $x_t \in \R^n$ be a stationary, mean-zero vector series observed on a
regular grid with spacing $\Delta$. Write
\[
S \;:=\; \E[x_t x_t'], \qquad
C(h) \;:=\; \E[x_{t+h}\, x_t'],
\]
so that $C(0) = S$. The contemporaneous covariance $S$ is symmetric by
construction. The lagged covariance $C(\Delta)$ is not, and its antisymmetric
part is the object of interest.

\begin{definition}[circulation]\label{def:circ}
The \emph{circulation} of $\{x_t\}$ at spacing $\Delta$ is
\begin{equation}\label{eq:qdelta}
Q_\Delta \;:=\; \tfrac{1}{2}\bigl(C(\Delta) - C(\Delta)'\bigr),
\qquad
(Q_\Delta)_{ij} = \tfrac12\bigl(\E[x_{i,t+\Delta} x_{j,t}]
- \E[x_{j,t+\Delta} x_{i,t}]\bigr).
\end{equation}
The \emph{net flow} of series $i$ is $\nu_i := 2\sum_j (Q_\Delta)_{ij}$.
\end{definition}

Three features of Definition~\ref{def:circ} deserve emphasis. It is
model-free: no linearity, Gaussianity, or Markov property is used, only
stationarity and finite second moments. It is antisymmetric by construction, so
it has $n(n-1)/2$ free entries and its diagonal is empty: a series has no
circulation with itself. And its sign has a fixed reading, namely
$(Q_\Delta)_{ij} > 0$ means that $j$ leads $i$, so that $j$ sends and $i$
receives; a positive net flow $\nu_i$ marks series $i$ as a net receiver, and
the net flows sum to zero across series.

For interpretation we use a benchmark. Let the data be generated by a
stationary Ornstein--Uhlenbeck system,
\begin{equation}\label{eq:ou}
dx_t = -A x_t\,dt + \Sigma^{1/2}\,dW_t,
\end{equation}
with $A$ having eigenvalues of positive real part, $\Sigma \succ 0$ (positive
definite, written $\succ 0$ throughout), and stationary covariance $S \succ 0$
solving the Lyapunov equation $A S + S A' = \Sigma$. Define
\begin{equation}\label{eq:q}
Q \;:=\; A S - \tfrac{1}{2}\Sigma .
\end{equation}
Substituting the Lyapunov equation gives $Q + Q' = 0$, so $Q$ is antisymmetric
and the drift decomposes as
\begin{equation}\label{eq:decomp}
A \;=\; \bigl(\tfrac{1}{2}\Sigma + Q\bigr)S^{-1}.
\end{equation}
The first term is self-adjoint in the geometry induced by $S$ and pulls the
system down the gradient of its stationary density, in plain terms the part of
the motion that runs downhill towards the long-run average. The second
transports probability mass around the level sets of that density without
changing it, the part that goes round and round. This is the finance and
econometrics reading of a decomposition that physics knows as the splitting of
a stationary diffusion into a detailed-balance part and a probability current
\citep{kwon2005structure,godreche2019characterising}; detailed balance, the
condition that every elementary flow is matched by its reverse, is the
classical probabilistic name for reversibility
\citep{kelly1979reversibility,jiang2004mathematical}. Sampling \eqref{eq:ou} at
spacing $\Delta$ gives the exact first-order vector autoregression
\begin{equation}\label{eq:var}
x_{t+\Delta} = \Phi\,x_t + \varepsilon_{t+\Delta}, \qquad
\Phi = e^{-A\Delta}, \qquad
\Sigma_\varepsilon = S - \Phi S \Phi',
\end{equation}
with $C(\Delta) = \Phi S$, and the two circulations are related by
$Q_\Delta = -\Delta\,Q + O(\Delta^2)$. The sign flip is a bookkeeping
consequence of \eqref{eq:q} and carries no content; throughout, the estimand is
the discrete object $Q_\Delta$, which needs no matrix logarithm and no
continuous-time interpretation to be computed.

\subsection{Identification}

\begin{assumption}[sampling bound]\label{ass:alias}
$\max_j \lvert \operatorname{Im} \lambda_j(A)\rvert < \pi/\Delta$.
\end{assumption}

\begin{theorem}[what identifies direction]\label{thm:ident}
Let the data be generated by \eqref{eq:ou} with $S, \Sigma \succ 0$.
\begin{enumerate}
\item[(i)] \emph{Contemporaneous covariance identifies no direction.} The pair
$(S, \Sigma)$ identifies only the gradient part $A_0 = \frac12\Sigma S^{-1}$.
The identified set for the drift is the affine space
\[
\mathcal{A}(S,\Sigma) = \Bigl\{\bigl(\tfrac12\Sigma + Q\bigr)S^{-1} :
Q' = -Q\Bigr\},
\]
of dimension $n(n-1)/2$; every member is automatically stable, and every
member has the same stationary law $N(0,S)$. Consequently every functional of the contemporaneous law, including
the correlation matrix, any distance transform of it, any graph built from such
distances, any undirected centrality, and the principal components, is constant
on $\mathcal{A}(S,\Sigma)$ and carries no information about $Q$.
\item[(ii)] \emph{The lag restores identification.} Under
Assumption~\ref{ass:alias}, $(S, C(\Delta))$ point-identifies $A$, and hence
$Q$, through $A = -\Delta^{-1}\log\bigl(C(\Delta)S^{-1}\bigr)$ with the
principal logarithm. The discrete circulation $Q_\Delta$ vanishes if and only
if $Q$ vanishes.
\item[(iii)] \emph{Direction is time asymmetry.} The following are equivalent:
(a) $Q = 0$; (b) $C(h) = C(h)'$ for every real $h \ge 0$; (c) the process is
statistically time reversible, $\{x_t\} \overset{d}{=} \{x_{-t}\}$. Under
Assumption~\ref{ass:alias} these are further equivalent to (d) $Q_\Delta = 0$
at the single sampled lag. Without the assumption, (d) is strictly weaker: a
system rotating by exactly a full turn per sampling interval has $Q \neq 0$
while every sampled lagged covariance is symmetric.
\end{enumerate}
\end{theorem}

Part (iii) for Gaussian processes goes back to \citet{weiss1975time}; the
identification condition in (ii) is that of \citet{phillips1973problem} and
\citet{hansen1983dimensionality}; the implications of observing only
$(S,\Sigma)$ in a Lyapunov model are studied by
\citet{varando2020graphical,dettling2023identifiability}. Part (i) is what
makes the assembly useful for applied work, and it has an immediate corollary.

\begin{corollary}[correlation networks and spanning trees]\label{cor:mst}
Any two systems with drifts $A_1, A_2 \in \mathcal{A}(S,\Sigma)$ generate
identical correlation matrices, identical Mantegna distances
$d_{ij} = \sqrt{2(1-\rho_{ij})}$, identical minimum spanning trees, and
identical undirected centralities, at every sampling of the same point-in-time
record, while their transmission directions may differ arbitrarily.
Correlations of temporally \emph{aggregated} records, sums of consecutive
observations rather than coarser samples of the same series, are not covered:
they involve the symmetric parts of $C(h)$ at $h \ge 1$, which do depend on
$Q$, and Section~\ref{sec:sampling} computes the size of that leak.
\end{corollary}

Corollary~\ref{cor:mst} is not an asymptotic statement and cannot be repaired
by more data. Section~\ref{sec:comparison} exhibits a pair of systems for which
the correlation matrices agree to $10^{\labCorrGapOrder}$ and the spanning
trees are \labMstIdentical, while one has no direction at all and the other
circulates.

A second warning is needed about the tools that do use lags. It is tempting to
expect that reversibility forces the pairwise transfer entropies, or
equivalently the pairwise Granger causalities \citep{barnett2009granger}, to be
symmetric. It does not.

\begin{remark}[predictive asymmetry is not direction]\label{rem:te}
Take $n = 2$ with unit variances, correlation $0.3$, and the symmetric lagged
covariance whose diagonal, the lag-one autocorrelation of each series, is
$0.5$ and $0.2$. The implied transition
matrix has real eigenvalues $\cexEigOne$ and $\cexEigTwo$ in $(0,1)$, so a
stationary sampled system with these moments exists; $C(\Delta)$ is symmetric,
so $Q_\Delta = 0$ and the system is exactly reversible. Yet the two Gaussian
transfer entropies are $\cexTeOne \times 10^{-3}$ and $\cexTeTwo \times
10^{-3}$ nats, differing by $\cexTeGap \times 10^{-4}$. The reason is visible
in the conditioning sets: $T_{j \to i}$ conditions on the past of $i$ and
$T_{i \to j}$ on the past of $j$, and when own persistences differ the two
conditionings are not exchangeable. Predictive asymmetry therefore mixes
circulation with heterogeneity in persistence. Under reversibility the mixture
is pure confound.
\end{remark}

\subsection{The magnitude of the arrow}

Reversibility is a point hypothesis and practice needs a size. The canonical
choice is the rate at which the forward law separates from the reversed law.

\begin{proposition}[entropy production]\label{prop:ep}
For the stationary system \eqref{eq:ou} the entropy production rate
\[
\sigma \;:=\; \lim_{T \to \infty} \tfrac{1}{T}
\KL\bigl(\mathbb{P}_{[0,T]} \,\Vert\, \mathbb{P}^{\mathrm{rev}}_{[0,T]}\bigr)
\;=\; 2\,\tr\bigl(Q'\Sigma^{-1} Q S^{-1}\bigr) \;\ge\; 0,
\]
with equality if and only if $Q = 0$. For the sampled chain \eqref{eq:var} the
per-period divergence $\sigma_\Delta$ between the forward and reversed
transitions has the exact expression \eqref{eq:exactkl} of
Appendix~\ref{app:proofs} and the two-term expansion
\begin{equation}\label{eq:sigq}
\sigma_\Delta = \sigma_Q + \sigma_C + O(\lVert Q_\Delta\rVert^3),
\qquad
\sigma_Q := 2\,\tr\bigl(Q_\Delta' \Sigma_\varepsilon^{-1} Q_\Delta S^{-1}\bigr),
\end{equation}
where $\sigma_C = \frac14\tr(E^2) \ge 0$ with
$E = \Sigma_\varepsilon^{-1}(\widetilde\Sigma - \Sigma_\varepsilon)$ and
$\widetilde\Sigma$ the reversed chain's innovation covariance. Both terms are
quadratic in $Q_\Delta$ to leading order and both vanish at $Q_\Delta = 0$,
but only $\sigma_Q$
carries the equivalence: $\sigma_Q = 0$ if and only if $Q_\Delta = 0$, whereas
$\sigma_C$ can vanish at nonzero circulation.
\end{proposition}

We report $\sigma_Q$ throughout. It is an exact quadratic functional of the
identified circulation, which is what makes unbiased estimation possible
(Section~\ref{sec:estimation}), it is nonnegative, and since $\sigma_C \ge 0$
it is a conservative reading of the full divergence rate to second order in
the circulation.

The unit is nats per period. The interpretation is the one that goes with a
Kullback--Leibler rate, read on the component that $\sigma_Q$ captures:
$\sigma_Q$ is the quadratic part of the expected rate at which the
log-likelihood ratio between the forward and the reversed description
accumulates, and since $\sigma_C \ge 0$ it understates that rate. A value of
$0.7$ nats per day therefore means that a day of data shifts the odds, on
average, by at least a factor of about two in favour of the forward reading. This gives a scale that
does not depend on how the series are normalised, and it produces a usable
trichotomy: $\sigma_Q = 0$ means no direction exists at second order, and the
feature ladder of the next subsection must be climbed before concluding that
none exists at all; $\sigma_Q$ small and positive means direction exists but
carries little information; $\sigma_Q$ large means directional structure is a
first-order feature of the joint law. One caution on comparisons: $\sigma_Q$
aggregates the contributions of all $n(n-1)/2$ pairs, so it grows with the
size of the panel at a fixed strength of circulation per pair. Comparisons
across cells of the same panel are clean; comparisons across panels of
different dimension should be made per pair, dividing by $n(n-1)/2$.

\begin{remark}[what the magnitude means away from the benchmark]\label{rem:bench}
$Q_\Delta$ is defined for any stationary series. For the Gaussian Markov
system matched to $(S, C(\Delta))$, $\sigma_Q$ is the quadratic
entropy-production component of the exact per-period divergence rate
$\sigma_\Delta$, and falls short of it by $\sigma_C \ge 0$; only in the
continuous-time limit do the two coincide. For data that are non-Gaussian or not Markov at the sampling
frequency, $\sigma_Q$ remains a well-defined, scale-free summary of
second-order irreversibility, and should be described as such; it is not in
general a bound on the true entropy production of the data-generating process,
in either direction. Higher-order irreversibility is reached through
Proposition~\ref{prop:feature} below.
\end{remark}

\subsection{Feature circulations}

The circulation of Definition~\ref{def:circ} reads only second moments. A
system can be strongly and unambiguously directional with every lagged
covariance equal to zero. The canonical example, used later as a test case, is
$y_t = c\,(x_{t-1}^2 - 1) + e_t$ with $x_t$ independent standard normal draws:
$y$ is driven entirely by the previous value of $x$ and by nothing else, yet
$\E[y_{t+1}x_t] = c\,\E[x_t^3] = 0$ and every entry of $C(\Delta)$ vanishes.
The repair is immediate and costs nothing.

\begin{definition}[feature circulation]\label{def:feat}
For measurable $\psi, \varphi$ with finite second moments under the stationary
law, the feature circulation is
\[
Q^{\psi\varphi}_\Delta \;:=\; \tfrac12\bigl(
\E[\psi(x_{t+\Delta})\,\varphi(x_t)] -
\E[\varphi(x_{t+\Delta})\,\psi(x_t)]\bigr).
\]
\end{definition}

Stacking a panel with a set of transforms and applying
Definition~\ref{def:circ} to the enlarged panel computes all such circulations
at once; the linear circulation is the case in which $\psi$ and $\varphi$ are
coordinate projections.

\begin{proposition}[feature circulations characterise reversibility]
\label{prop:feature}
Let $\{x_t\}$ be stationary with stationary law $\pi$.
\begin{enumerate}
\item[(i)] If $\{x_t\}$ is reversible then $Q^{\psi\varphi}_h = 0$ for every
$h$ and every pair $\psi, \varphi \in L^2(\pi)$.
\item[(ii)] If in addition $\{x_t\}$ is Markov, the converse holds: if
$Q^{\psi\varphi}_\Delta = 0$ for all $\psi, \varphi$ in a complete system of
$L^2(\pi)$, then the process is reversible.
\end{enumerate}
\end{proposition}

Part (ii) is the statement that reversibility of a stationary Markov process is
exactly self-adjointness of its transition operator on $L^2(\pi)$, and that
self-adjointness can be checked on any complete system. The practical
consequence is a ladder: the linear circulation is the first rung, adding
squares reaches quadratic transmission of the kind in the example above, and
richer bases reach further, at the cost of estimating more entries from the
same data. Every rung is tested with the same exact reference distribution,
since Section~\ref{sec:inference} constructs it from the null hypothesis and
not from the statistic.

Two cautions. First, feature circulations of a single series with its own
transforms, for instance $x_i$ against $x_i^2$, detect asymmetries within one
series, such as the tendency of volatility to rise faster than it falls. Those
are real irreversibilities, but they are not cross-series transmission, and an
analyst asking about transmission should restrict the reading to entries
linking two different underlying series. Section~\ref{sec:inference} reports
both. Second, adding transforms multiplies the number of entries and therefore
the noise; the bias correction of the next section becomes more important, not
less, as the basis grows.

\subsection{What the framework identifies}\label{sec:identifies}

Table~\ref{tab:identifies} summarises the section by object. Each row names a
quantity, the question it answers, and the question it is sometimes read as
answering but does not. The table encodes the principle that organises the
rest of the paper: irreversibility, predictability, structural causality, and
counterfactual transmission are four different properties of a system, and no
one of them implies the next. Sections~\ref{sec:comparison} and
\ref{sec:design} give each of these distinctions an operational form.

\begin{table}[t]
\centering
\caption{What each object answers, and what it does not.}
\label{tab:identifies}
\small
\begin{tabular}{>{\raggedright\arraybackslash}p{0.30\textwidth}>{\raggedright\arraybackslash}p{0.29\textwidth}>{\raggedright\arraybackslash}p{0.32\textwidth}}
\toprule
Object & Answers & Does not answer \\
\midrule
Contemporaneous covariance $S$ & which series move together & direction, by
Theorem~\ref{thm:ident}(i) \\[2pt]
Predictive asymmetries (Granger causality, transfer entropy) & who helps
forecast whom & irreversibility, by Remark~\ref{rem:te} \\[2pt]
Circulation $Q_\Delta$ & whether and where the law distinguishes forward from
backward time & structural causality \\[2pt]
Magnitude $\sigma_Q$ & how strongly the law distinguishes the two directions
& the effect of any intervention \\[2pt]
Structural model & mechanism, under identifying assumptions & model-free
direction \\
\bottomrule
\end{tabular}
\end{table}

\section{Estimation}\label{sec:estimation}

\subsection{Moment estimators}

Let the sample consist of one or more contiguous stretches
$e_1, \dots, e_E$ of the series, which we call episodes. Episodes matter because
applied work often wants a parameter within a regime rather than over the whole
record, and because no product of observations should ever straddle a gap
between two stretches. Within each episode the series are centred and scaled,
and the two moment matrices are
\begin{equation}\label{eq:moments}
\widehat{S} = \frac{1}{T}\sum_{t} x_t x_t', \qquad
\widehat{C}(\Delta) = \frac{1}{\sum_e (T_e - 1)}
\sum_{e}\sum_{t \in e,\, t+\Delta \in e} x_{t+\Delta}\, x_t',
\end{equation}
from which the circulation, the implied transition matrix, the innovation
covariance, and the magnitude follow:
\begin{equation}\label{eq:estimators}
\widehat{Q}_\Delta = \tfrac12\bigl(\widehat{C}(\Delta) - \widehat{C}(\Delta)'\bigr),
\quad
\widehat{\Phi} = \widehat{C}(\Delta)\widehat{S}^{-1},
\quad
\widehat{\Sigma}_\varepsilon = \widehat{S} - \widehat{\Phi}\widehat{S}\widehat{\Phi}',
\end{equation}
and $\widehat\sigma_Q = 2\tr(\widehat{Q}_\Delta'
\widehat{\Sigma}_\varepsilon^{-1}\widehat{Q}_\Delta \widehat{S}^{-1})$.
Algorithm~\ref{alg:core} of Appendix~\ref{app:algorithms} states this in
eight steps of plain arithmetic. There is no optimisation, no tuning parameter, and
no iterative step anywhere in the estimator.

\subsection{Bias in quadratic functionals}

The distinction that governs everything in this section is between readings of
$\widehat{Q}_\Delta$ that are linear in it and readings that are quadratic.

\begin{lemma}[bias]\label{lem:bias}
Under stationarity with known episode means, every entry of
$\widehat{Q}_\Delta$ is an unbiased estimator of the corresponding entry of
$Q_\Delta$, and so is every linear functional of it, including the net flows.
Estimating the episode means and scales contributes a bias that is symmetric
across the pair of indices to leading order and therefore cancels in the
antisymmetric part. The plug-in estimator of the quadratic functional
$\sigma_Q$ carries a positive bias of order $n^2/T$.
\end{lemma}

The mechanism is elementary and worth stating in words, because it is the
single most consequential fact for applied use. Write $\widehat{Q} = Q + \xi$
with $\E[\xi] = 0$. Then
$\E[\tr(\widehat{Q}'W\widehat{Q}V)] = \tr(Q'WQV) + \E[\tr(\xi'W\xi V)]$, and
the second term equals $\E\lVert W^{1/2}\xi V^{1/2}\rVert_F^2 \ge 0$, an
accumulation of the estimation error's second moments, each of order $1/T$,
over the $n(n-1)/2$ free entries. The estimator therefore
reports the noise as if it were signal, and it does so with a magnitude that
grows with the square of the number of series. Table~\ref{tab:grid} measures
this in systems whose true magnitude is exactly zero.

\begin{table}[t]
\centering
\caption{The plug-in estimate of the magnitude when the truth is zero, against
the cross-fitted estimate. Mean over $\gridM$ replications of a stationary
reversible system.}
\label{tab:grid}
\small
\begin{tabular}{rrrrr}
\toprule
$n$ & $T$ & $n^2/T$ & plug-in & cross-fitted \\
\midrule
5 & 200 & 0.125 & +0.113 & +0.005 \\
5 & 400 & 0.062 & +0.048 & -0.006 \\
5 & 800 & 0.031 & +0.025 & +0.002 \\
5 & 1600 & 0.016 & +0.012 & +0.000 \\
10 & 200 & 0.500 & +0.518 & +0.025 \\
10 & 400 & 0.250 & +0.234 & -0.003 \\
10 & 800 & 0.125 & +0.112 & +0.000 \\
10 & 1600 & 0.062 & +0.054 & -0.002 \\
26 & 200 & 3.380 & +4.413 & +0.280 \\
26 & 400 & 1.690 & +1.853 & +0.051 \\
26 & 800 & 0.845 & +0.845 & -0.009 \\
26 & 1600 & 0.422 & +0.405 & -0.008 \\
\bottomrule
\end{tabular}

\begin{minipage}{0.86\textwidth}
\vspace{1ex}\footnotesize
Notes. Both columns report the mean estimate of $\sigma_Q$ in nats per period
when the data-generating process is exactly reversible, so that the correct
answer is zero in every row. Across the grid the ratio of the plug-in mean to
$n^2/T$ lies between $\gridRatioMin$ and $\gridRatioMax$: the rate is the
content of Lemma~\ref{lem:bias}, the constant depends on the weights.
\end{minipage}
\end{table}

At twenty-six series and eight hundred observations, dimensions typical of a
panel of national markets over one crisis, the plug-in estimator reports
$\gridPluginBig$ nats per day when the truth is zero, against a ratio
$n^2/T = \gridRatioBig$. At five series and the same length it reports
$\gridPluginSmall$. Any comparison of magnitudes across panels of different
size, or across regimes of different length, is contaminated at first order by
this term. It is not a small-sample correction; in the applications we have in
mind it is the entire measured quantity.

\subsection{Cross-fitting}

The remedy follows from the algebra. The bias arises because the same
estimation error appears in both factors of the quadratic. Estimate the two
factors on disjoint stretches of time and the cross term has expectation zero.
The device of splitting a sample so that two estimation errors cannot multiply
is the same one that removes the first-order bias in semiparametric estimation
\citep{chernozhukov2018double}; here the split is along time rather than across
observations, because time is where the dependence lives.

\begin{definition}[cross-fitted magnitude]\label{def:cf}
Within each episode assign the lagged pairs to two folds by alternating blocks
of $L_f$ observations. Let $\widehat{Q}_A$ and $\widehat{Q}_B$ be the
circulations computed on the two folds. Then
\begin{equation}\label{eq:cf}
\widehat\sigma_Q^{\,\mathrm{cf}} \;=\;
\tr\bigl(\widehat{Q}_A' \widehat{\Sigma}_\varepsilon^{-1}
\widehat{Q}_B \widehat{S}^{-1}\bigr)
+ \tr\bigl(\widehat{Q}_B' \widehat{\Sigma}_\varepsilon^{-1}
\widehat{Q}_A \widehat{S}^{-1}\bigr),
\end{equation}
with the weights estimated on the full sample.
\end{definition}

Three points about \eqref{eq:cf}. The two folds are dependent only through
observations adjacent to block boundaries, so with $L_f$ large relative to the
dependence range the residual bias is of smaller order; we use
$L_f = 50$ observations throughout and report the sensitivity in
Appendix~\ref{app:estimation}. The weights $(\widehat{\Sigma}_\varepsilon^{-1},
\widehat{S}^{-1})$ are estimated on the whole sample and their error enters the
level of a functional that is already quadratic in the small quantity
$Q_\Delta$, so it contributes at second order. And the estimator can be
negative in finite samples, which is a feature rather than a defect: an
estimator of a nonnegative quantity that is never negative cannot be unbiased
at the boundary, and truncating at zero would reintroduce exactly the upward
bias that cross-fitting removes.

The right column of Table~\ref{tab:grid} shows the result: at zero truth the
cross-fitted estimator returns $\gridCfBig$ where the plug-in returns
$\gridPluginBig$, and it stays near zero across the grid. The one cell where it
degrades is the extreme case $n = 26$, $T = 200$, where $n^2/T$ exceeds three
and the cross-fitted mean is $\gridCfWorst$ against a plug-in
$\gridPluginWorst$: cross-fitting reduces the problem by an order of magnitude
but does not license estimation at arbitrary dimension.

Away from zero the picture is the same. In a system with true magnitude
$\covTrue$ nats per period at $n = 26$ and $T = 800$, over $\covM$
replications, the cross-fitted estimator averages $\covCf$, a bias of
$\covCfBias$, while the plug-in averages $\covPlugin$: an inflation of the
reported magnitude by a factor of $\covInflate$. The small residual bias of
the cross-fitted estimator, $\covCfRelBias$ per cent here, has a known source:
cross-fitting removes the additive term of Lemma~\ref{lem:bias}, but the
weights $(\widehat\Sigma_\varepsilon^{-1}, \widehat S^{-1})$ are still
estimated, and an estimated inverse covariance is upward biased by a factor of
order $T/(T-n)$. That multiplicative term is invisible at zero truth, which is
why Table~\ref{tab:grid} cannot show it, and it cancels in any ratio or
difference of magnitudes computed with shared weights.

\subsection{Block jackknife standard errors}

The natural first thought for standard errors is a moving-block bootstrap.
It fails here, and the reason is instructive. A bootstrap sample drawn with
replacement places the same original observations into both cross-fit folds, so
the very dependence that cross-fitting was designed to break is restored inside
each bootstrap replicate, and the bootstrap distribution recentres on the
biased plug-in. The symptom is unmistakable when it happens: confidence
intervals that do not contain their own point estimate.

We therefore use a delete-one-block jackknife, the deletion analogue of the
block methods for dependent data \citep{kunsch1989jackknife}. The sample is
partitioned into blocks of roughly $L_j$ observations within episodes, the
statistic is recomputed with each block deleted in turn, and
\[
\widehat{\mathrm{se}}^{\,2}
= \frac{K-1}{K}\sum_{k=1}^{K}\bigl(\widehat\theta_{(-k)} - \bar\theta\bigr)^2 .
\]
No observation appears twice in any replicate, so the fold structure survives;
Algorithm~\ref{alg:jack} states the loop.
In the calibration above the resulting intervals cover the truth in $\covJk$ of
replications against a nominal $0.95$, with the shortfall attributable to the
small number of blocks available within an episode. Coverage, block-length
sensitivity, and the behaviour of the discarded bootstrap are documented in
Appendix~\ref{app:estimation}.

\subsection{Implementation choices}

The estimator has few free choices, and the following defaults have been used
in every application so far.

\emph{Episodes.} Define them by an external or estimated criterion, never by
the statistic itself, and never let a lagged product cross a boundary. If
regimes are estimated, the sensitivity of the conclusions to the regime dating
must be reported; Section~\ref{sec:design} gives the two checks we use.

\emph{Standardisation.} Scale within episode, not over the whole record.
Scaling across regimes of different volatility imports the volatility
difference into the circulation.

\emph{Fold and block lengths.} Fold blocks of $50$ observations and jackknife
blocks of $100$ are the defaults; both should be long relative to the
dependence range of the series and short enough to leave a usable number of
blocks. Since the two serve different purposes, they need not agree.

\emph{Time scales.} Many applications care about the horizon at which
transmission operates. A wavelet decomposition of each series into scale
components, followed by the same estimator applied to each component, gives a
scale-by-scale reading, and the exact test of Section~\ref{sec:inference}
remains valid provided the reversal is applied before the filter. The paper's
application uses four scales; nothing in the method requires any.

\emph{Winsorisation.} Extreme observations enter the second moments with
squared weight. We winsorise at the $0.1$ and $99.9$ percentiles of the full
sample before anything else. Because the same transformation is applied to the
reversed records under the null, this does not disturb the exactness of the
test.

\section{Inference}\label{sec:inference}

\subsection{An exact randomisation test}

The hypothesis of no direction is a symmetry: the record and its reverse have
the same law. Symmetries are the ideal setting for randomisation inference,
because the transformations that leave the null law invariant can be applied to
the observed data to manufacture as many equally likely records as required.
Within the stated block structure of the null, no asymptotic approximation,
no variance estimate, and no distributional assumption enters anywhere.

\begin{proposition}[exactness of the reversal test]\label{prop:reversal}
Partition each episode into blocks $B_1, \dots, B_K$ of length at least $L$, and
let $g = (g_1, \dots, g_K) \in \{0,1\}^K$ act on the observed record by
reversing time within block $k$ whenever $g_k = 1$. Under the null hypothesis
that the within-episode law is reversible and the blocks are mutually
independent, $\{g \cdot x\}$ is an exchangeable family, and for any statistic
$T$ the randomisation $p$-value
\[
p \;=\; \frac{1 + \#\{r : T(g_r \cdot x) \ge T(x)\}}{R+1}
\]
over $R$ uniform draws satisfies $\Pr(p \le \alpha) \le \alpha$ at every sample
size. Exactness holds as stated when dependence does not cross block
boundaries; when it does, with range $m$, the perturbation to the statistic is
of order $O(m/L)$, exactness holds for the record with the block boundaries
excised, and we treat the un-excised test as an approximation at that order.
\end{proposition}

Three practical remarks follow. First, the choice of statistic is free:
exactness is a property of the group
action, not of $T$. The analyst may use the circulation norm
$\lVert \widehat{Q}_\Delta\rVert_F$, the cross-fitted magnitude
$\widehat\sigma_Q^{\,\mathrm{cf}}$, a single entry, a net flow, or any feature
circulation of Section~\ref{sec:framework}, and the same draws serve all of
them. This is what makes the nonlinear extension cheap: a richer statistic
changes the power of the test and nothing about its validity.

Second, resolution is set by the number of draws. The smallest attainable
$p$-value is $1/(R+1)$, so $R$ should be chosen so that $1/(R+1)$ sits
comfortably below the significance level divided by the number of tests. The
studies in this paper use $R$ between $299$, for simulation studies repeated
hundreds of times, and $2{,}000$, for the tests that carry inferential weight.

Third, the transformation must be applied to the rawest available record.
This is the trap that costs the most, and it deserves its own subsection.

\subsection{Filtering and the order of operations}

Suppose the series are filtered before estimation, for instance decomposed into
time scales by a wavelet transform, or smoothed, or passed through any linear
filter with an asymmetric impulse response. Nearly every filter in practical
use is asymmetric in this sense. Such a filter manufactures time
irreversibility: the filtered output of a perfectly reversible input is not
reversible, because the filter itself points forwards in time.

A test that reverses the filtered coefficients is therefore testing a null
under which the filter's own asymmetry is absent, which is not the hypothesis
of interest, and the test rejects far too often. The repair is to apply the
reversal to the raw record and then to recompute everything, filtering included,
on the reversed record. The filter's contribution then appears identically
under the null and cancels. Algorithm~\ref{alg:reversal} states the loop.

The same principle governs every other preprocessing step, with one exact
exemption. A step that depends on the sample only through quantities invariant
under reversal, winsorisation thresholds and full-sample scale factors are the
common cases since reordering observations does not change them, may be
computed once outside the loop with nothing lost. Every step that is not
invariant, filtering and scale decomposition above all, must be recomputed on
each reversed record, because any non-invariant step held fixed contributes to
the statistic something that is absent from the reference distribution. An
estimated regime dating sits in between: it is not invariant, and if it is
held fixed the test is exact only conditionally on the dating. The right
course is to say so and to check the conditioning, which is what the
end-to-end placebo of Section~\ref{sec:design} does: reversible input with
realistic volatility is passed through the full construction, the empirical
episode structure included, and the rejection rate stays at the nominal
level.

\subsection{Multiple testing}

Applications rarely ask one question. A panel decomposed into two regimes and
four time scales produces eight cells, and reporting the smallest of eight
$p$-values without adjustment is the oldest mistake in the book. The
randomisation scheme provides the correction for free: applying the same draws
$g_1, \dots, g_R$ to every cell yields draws from the joint null distribution
of the vector of cell statistics, which is precisely the input required by the
stepdown method of \citet{romano2005stepwise}. Familywise error is then
controlled without a single additional simulation and without assuming
independence across cells, which would be false here. Algorithm~\ref{alg:rw}
gives the stepdown loop.

\subsection{Conditional tests}

Many designs ask a conditional question: whether a sending series predicts
the receiver \emph{given} a set of controls. The natural randomisation
analogue is
to rotate the sender in time, which preserves its marginal distribution and
its autocorrelation while destroying its alignment with the receiver. One
caveat before the construction: unlike block reversal under the
reversibility null, a cyclic rotation does not leave the law of a finite
stretch of a stationary series exactly invariant, because the rotation joins
the end of the record to its beginning. The test below is therefore valid up
to a wrap-around error that vanishes with the sample length, of order $1/T$,
rather than exactly; at the sample sizes of the application the distinction is
immaterial, but it should not be stated away.

The detail that matters is what gets rotated. Rotating the raw sender destroys
not only its alignment with the receiver, which the null requires, but also its
dependence on the conditioning variables, which the null does not touch. The
resulting reference distribution corresponds to a stronger null than the one
being tested, and the test is anti-conservative. The repair is the
residualisation device introduced for permutation tests by
\citet{freedman1983nearly}, in the variant of \citet{kennedy1995randomization}:
residualise both the receiver and the sender on the conditioning set once, and
rotate the residualised sender.

\begin{algo}[Conditional circular-shift test]\label{alg:shift}
Given receiver $y$, sender $u$, controls $Z$ (including an intercept), and $R$
draws:
\begin{enumerate}\itemsep2pt
\item $r_y \leftarrow y - Z(Z'Z)^{-1}Z'y$ and $r_u \leftarrow u -
Z(Z'Z)^{-1}Z'u$.
\item Observed statistic: $\hat\rho \leftarrow \operatorname{corr}(r_y, r_u)$,
reported in information units as $\widehat{T} = -\frac12\log(1-\hat\rho^2)$.
\item For $r = 1, \dots, R$: draw a shift $s_r$ uniformly from
$\{L_s, \dots, T-L_s\}$, where $T$ is the length of the series and $L_s$ a
small guard, $30$ in the reference implementation, that keeps every rotation
far enough from the identity; form the rotation $r_u^{(s_r)}$ and recompute
the statistic.
\item Report $p = (1 + \#\{r : T^{(r)} \ge \widehat T\})/(R+1)$.
\end{enumerate}
\end{algo}

The two constructions can differ materially. In the application of
Section~\ref{sec:application} the corrected test returns
$p = \appPShiftCalm$ for a channel where rotating the raw sender returned a
$p$-value roughly three times smaller.

\subsection{Power and size}

\begin{figure}[t]
\centering
\includegraphics[width=\textwidth]{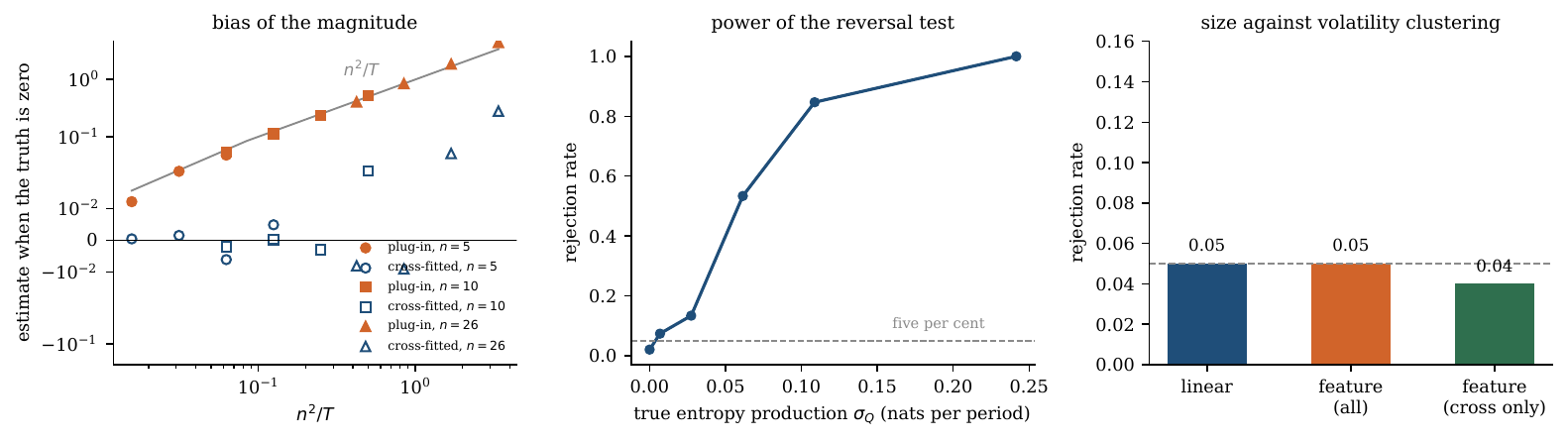}
\caption{Estimator and test behaviour. Left: the mean estimate of $\sigma_Q$
when the truth is zero, against $n^2/T$, for the plug-in (filled) and
cross-fitted (open) estimators, with the line $n^2/T$ for reference. Centre:
rejection rate of the reversal test at the five per cent level against the true
magnitude, $n = 10$, $T = 800$, $\powM$ replications. Right: rejection rate
against a reversible process with volatility clustering, for the linear
statistic and for the two feature-map statistics, $\szM$ replications.}
\label{fig:estimator}
\end{figure}

Validity is cheap; power is not. Figure~\ref{fig:estimator} reports both, from
systems where the answer is known.

\emph{Power.} At $n = 10$ and $T = 800$ the reversal test rejects in
$\powZero$ of samples when the truth is reversible, rising to $\powMid$ at
$\sigma_Q = \powMidSig$ and $\powTop$ at $\sigma_Q = \powTopSig$ nats per
period. The conservativeness at zero is expected: the randomisation
$p$-value is valid but not exactly uniform when the statistic has a discrete
null distribution and the block structure is coarse.

\emph{Size against the obvious nuisance.} Financial and biological series
display volatility clustering, and the concern is that clustering alone might
be read as an arrow. It is not: against a reversible process with fitted
volatility dynamics the reversal test rejects in $\szLinear$ of samples using
the linear statistic, $\szFeatAll$ using the full feature statistic, and
$\szFeatCross$ using the cross-series feature statistic, all against a nominal
five per cent. Volatility clustering by itself does not manufacture a
rejection.

\emph{The blind spot.} Against the purely quadratic
transmission described in Section~\ref{sec:framework} the linear statistic
rejects in $\labRevD$ of samples, the nominal rate and no more, because the
population value of every lagged covariance is exactly zero and the test has
nothing to detect. The feature statistic
rejects in $\labFeatD$. An analyst who has reason to expect nonlinear
transmission should climb the ladder rather than conclude from a linear null
result that nothing is there.

\section{Sampling Theory}\label{sec:sampling}

\begin{figure}[t]
\centering
\includegraphics[width=\textwidth]{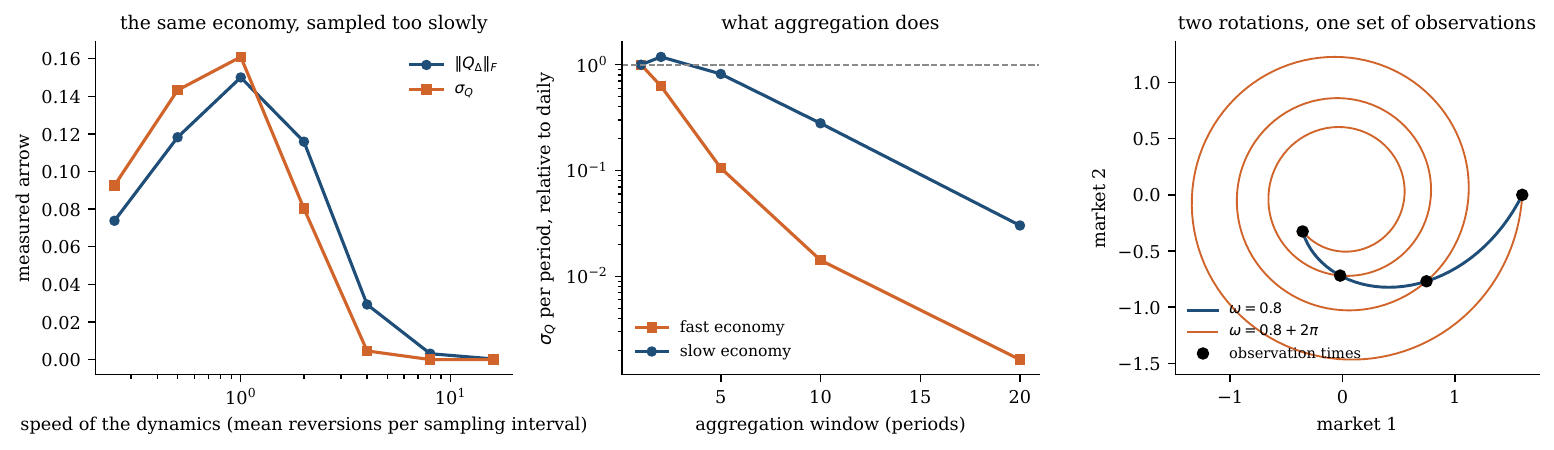}
\caption{Sampling and the measured arrow. Left: the same system, with the
stationary law held fixed, made progressively faster relative to the
observation interval; the measured circulation and magnitude rise, peak, and
collapse. Centre: the magnitude per period of non-overlapping sums, relative to
the value at the original frequency, for a fast and a slow system. Right: two
continuous paths whose rotation frequencies differ by exactly one turn per
observation interval; they pass through the same observations (dots) and no
record on that grid can separate them.}
\label{fig:sampling}
\end{figure}

A record is a sequence of snapshots. Whether an arrow of time can be seen in it
depends on how the interval between snapshots compares with the speed of the
system, and the dependence is severe enough that an analyst should establish it
before estimating anything. This section states the three effects, each with a
number attached. Figure~\ref{fig:sampling} displays them.

\subsection{Aliasing}

Assumption~\ref{ass:alias} is not a technical convenience. If a system rotates
faster than half a turn per observation interval, the record it leaves is
identical to that of a slower system. The right panel of
Figure~\ref{fig:sampling} shows an explicit pair: two systems with the same
decay and rotation frequencies $\aliasOmegaOne$ and $\aliasOmegaTwo$ radians
per unit time, differing by exactly $2\pi$ per interval. Their continuous
circulations are $\aliasQOne$ and $\aliasQTwo$, differing by a factor of nearly
nine, yet their sampled transition matrices agree to
$10^{\aliasPhiGapOrder}$ and both produce the identical discrete circulation
$\aliasQDelta$. No estimator, no sample size, and no cleverness separates them.
What the record does identify, uniquely, is the member of the family whose
rotation is slower than half a turn per interval, recovered here by the
principal logarithm to $10^{\aliasRecoverOrder}$.

The practical reading is not despair but modesty about the target. The discrete
circulation $Q_\Delta$ is always identified and always interpretable: it is the
net directed flow at the frequency observed. The continuous-time generator
behind it is identified only under Assumption~\ref{ass:alias}. Since applied
work almost always wants the former, this paper reports the former.

\subsection{The small-interval approximation}

For intervals short relative to the system's own dynamics the relation between
the discrete and continuous objects is simple: $Q_\Delta = -\Delta\,Q +
O(\Delta^2)$. Over intervals from $0.02$ to $1$ mean-reversion times the
relative error of the linear approximation grows with slope
$\smallIntSlope$ in logarithms, confirming that the neglected term is second
order. For an analyst this says that at high frequency the measured circulation
scales proportionally with the interval, so that comparisons across sampling
frequencies require nothing more than a rescaling. At lower frequency it does
not, which is the subject of the next two subsections.

\subsection{The window of visibility}

Hold the stationary distribution of a system fixed and change only its speed.
The correlation matrix does not move at all, by construction, to
$10^{\tsCorrDevOrder}$. The measured arrow, however, traces an inverted U. In
the left panel of Figure~\ref{fig:sampling}, a system with a half-life of
$2.8$ observation intervals shows circulation $\tsSlowQ$; the peak is at a
half-life of about $0.7$ intervals with circulation $\tsPeakQ$; and a system
$\tsFastKappa$ times faster than that, with a half-life of about a twentieth of
an interval, shows $\tsFastQ$, which is to say nothing.

Both ends have intuitive readings. A system far slower than the observation
interval has barely moved between snapshots, so the directed part of that
movement is small. A system far faster has completed its circulation and
returned to its stationary distribution before the next snapshot, so the
snapshots are effectively independent draws from a law that carries no
directional information. Direction is visible in the middle, when the system
moves appreciably but not completely between observations.

This is a design constraint, not an estimation problem. An analyst who suspects
transmission at a horizon of minutes will not find it in daily data no matter
how many years are collected. The corresponding diagnostic is cheap: fit the
transition matrix, read off the implied half-lives, and check that they are of
the same order as the observation interval. We recommend reporting them
alongside any circulation estimate.

\subsection{Aggregation}

Suppose the data are available at one frequency and the analyst reports results
at a coarser one, by summing returns over weeks rather than days. What happens
to the arrow is a parameter-free prediction of the fitted model, because the
lagged covariance of non-overlapping sums is a fixed linear combination of the
underlying autocovariances,
\[
C_Y(1) \;=\; \sum_{a=1}^{w}\sum_{b=1}^{w} C\bigl((w + a - b)\Delta\bigr),
\]
and $C(h) = \Phi^h S$ for the benchmark. The centre panel of
Figure~\ref{fig:sampling} evaluates this. For a fast system, aggregation
destroys the arrow: the magnitude per period falls to $\aggFastTwo$ of its
original value at two periods, $\aggFastFive$ at five, and $\aggFastTen$ at
ten. For a slow system, moderate aggregation \emph{amplifies} it, to
$\aggSlowTwo$ at two periods, before the same eventual decay sets in
($\aggSlowFive$ at five, $\aggSlowTen$ at ten).

Two consequences. First, the apparent disappearance of a directional result
under aggregation is not by itself evidence that the original result was
spurious; the size of the expected attenuation is computable from the fitted
model, and the informative comparison is between the observed change and the
predicted one. Second, since the prediction is parameter-free, the comparison
is a genuine out-of-sample check on the fitted model, and we recommend running
it, prepared for it to fail informatively: in the application, the observed
weekly-to-daily attenuation at the finest crisis scale is $\attObs$ against a
predicted $\attPred$, so the arrow survives aggregation better than the fitted
linear benchmark says it should, a discrepancy the companion study reports and
discusses. In the systems here, where the benchmark is true by construction,
the analytic aggregation formula reproduces the simulated moments with a
largest deviation of $\aggCheckT$ standard errors across $\aggCheckK$
independent replicates, which is what sampling noise alone would produce.

Aggregation also resolves a loose end left by Theorem~\ref{thm:ident}(i). The
theorem protects the contemporaneous covariance of the same point-in-time
record at every sampling; sums are not the same record. Because the variance
of a $w$-period sum involves the symmetric parts of $C(h)$ at lags $h \ge 1$,
which do depend on the circulation, the reversible and circulating twins of
Section~\ref{sec:comparison}, identical in correlation at every sampling,
separate slightly once aggregated: the largest absolute difference between
their correlation matrices of non-overlapping sums is $\aggLeakTwo$ at
$w = 2$, $\aggLeakFive$ at $w = 5$, and $\aggLeakTen$ at $w = 10$. The leak is
real, computable, and an order of magnitude too small to rescue the
correlation toolkit: it identifies the presence of some asymmetry, at some
aggregation, without recovering its pattern.

\section{Research Design}\label{sec:design}

\subsection{The gap between measurement and transmission}

Everything so far concerns a property of the observed law. Whether that
property licenses a claim about transmission is a separate question, and no
estimator settles it. The cleanest way to see the gap is to construct a system
that has a large, real, correctly measured arrow of time and no transmission of
any kind.

Let a single stream of news arrive continuously, and let three markets record a
closing price at three different hours of the day. Each market's daily return
accumulates the news of the twenty-four hours ending at its own close. Nothing
passes from one market to another; there is one common driver and three
observation times. Yet the market that closes last has already recorded news
that the market closing first will only record on the following day, so the
lagged covariance is asymmetric and the circulation is nonzero. In the
calibration used later this system has circulation $\labQfC$ and magnitude
$\labSigC$ nats per period, larger than in the system where genuine circulating
dynamics were deliberately installed. Every directional method in the
comparison, including ours, reports a confident direction. Every one of them is
right about the arrow and wrong about transmission, if transmission is what is
being claimed.

This is not a defect that better statistics can repair. Both readings, the
mechanical and the economic, are true statements about the sampled law. They
are separated only by a research design that finds an observable point where
they disagree.

\subsection{A separating design}

\begin{figure}[t]
\centering
\includegraphics[width=\textwidth]{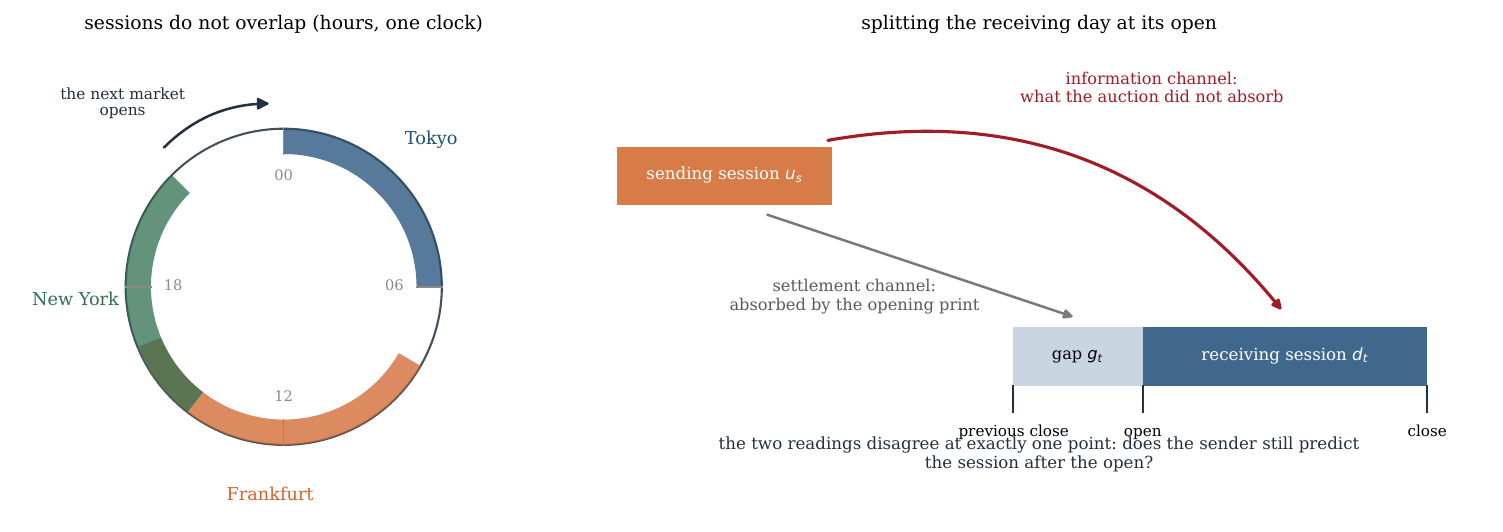}
\caption{The design. Left: sessions in different time zones do not overlap, so
the trading clock orders them mechanically. Right: the receiving day is split
at its opening auction into an overnight gap and an intraday session, and the
sending session is tested against each component separately.}
\label{fig:ladder}
\end{figure}

The design we use exploits the fact that mechanical ordering and economic
transmission make different predictions about a market's opening auction.
Figure~\ref{fig:ladder} sets it out. If the arrow from an earlier session into a
later one is nothing but the ordering of clocks, then the receiving market's
opening auction, which takes place after the sending session has closed and
with its outcome public, prices that information completely, and nothing of the
sending session predicts the receiving market's subsequent trading. If instead
information continues to propagate after the open, the sending session predicts
the receiving session conditional on the open.

Formally, decompose the receiving market's day $t$ into the overnight gap $g_t$
(previous close to open) and the intraday session $d_t$ (open to close), and
define two channels with respect to a sending session return $u_s$ that ends
strictly before the receiver's open:
\begin{align*}
\text{settlement channel:}\quad & g_t \sim u_s \mid g_{t-1}, d_{t-1},\\
\text{information channel:}\quad & d_t \sim u_s \mid g_t,\, g_{t-1}, d_{t-1}.
\end{align*}
The conditioning on $g_t$ in the second line is what does the work: it asks
what the sender adds beyond what the opening print already absorbed. Each
channel is summarised by a partial correlation converted to information units,
tested by the conditional randomisation of Algorithm~\ref{alg:shift}, and given
a heteroskedasticity and autocorrelation robust standard error.

The reading must be committed before estimation, and both outcomes must be
reportable. A positive information channel is transmission. A zero information
channel with a positive settlement channel is price discovery at the open, that
is, the mechanical reading. In the application both occur, in different
regimes and for different senders, which is only informative because the design
was set up to permit either.

\subsection{Auditing the design inputs}

A design is only as good as the observable it hinges on, and here that is the
opening price. Opening prints are the least reliable field in most market data
feeds: some are stale copies of the previous close, some are the first
transaction of the day at an arbitrary hour, and some are the auction price the
design assumes. A design that rests on them without checking is a design that
rests on a data vendor's conventions.

We therefore apply screens before estimation, per series and per year: the
share of days with open equal to close, and the share with open equal to the
\emph{previous} close, must both lie below fixed thresholds, and a series-year
that fails either is excluded. In the application these screens excluded one
major index outright and forced the replacement of another index by an
exchange-traded vehicle whose opening prints are genuine transactions. The
screens are stated in Appendix~\ref{app:estimation}; the point of general
interest is that they are pre-committed, mechanical, and applied without
reference to the results.

\subsection{Placebo tests}

A design should be run on cases whose answer is known. Three kinds have proved
worth the effort.

\emph{A timing placebo.} Repeat the estimation with the sender lagged by one
further period, so that its content was already public before the receiver's
previous close. Any genuine sequencing effect must vanish. In the application
this placebo annihilates channels whose original statistics exceeded ten
standard errors, which is the strongest available evidence that the design
measures session ordering and not persistence.

\emph{An end-to-end placebo of the whole procedure.} Simulate a reversible
process with realistic nuisance features, here volatility clustering and heavy
tails, and pass it through every step actually used, including the episode
structure of the estimated regime dating, the filtering, and the test. Rejection rates should equal the
nominal level. In the application they lie between $\appPlcMin$ and
$\appPlcMax$ over $\appPlcM$ replications at a nominal five per cent. This
catches errors that no individual component test would: it is the only check
that covers the interaction between regime selection and inference.

\emph{A discriminant against the obvious alternative.} If crisis regimes might
be proxying something else, that something else should be measured directly and
tested. In the application, an event study of a geopolitical risk index around
regime entries finds flat pre-trends and no post-entry elevation, which
separates the financial-stress reading from the geopolitical one.

\subsection{Attribution, anatomy, and the stability of the regime dating}

Three further steps convert a detected channel into an interpretable one.

\emph{Attribution.} When two candidate senders are correlated, a channel found
for one may belong to the other. Placing both in the same conditional
regression settles it. In the application this attributes the crisis channel to
the European session rather than to the correlated United States session.

\emph{Anatomy.} A channel may operate on the whole distribution or only in its
tail. Splitting the sender at a quantile and adding an interaction distinguishes
the two, and the answer changes the economic interpretation: a tail-only channel
points to a constraint that binds in large moves, not to a general
strengthening of linkages. In the application the crisis channel is carried by
the largest decile of sending shocks, with an interaction coefficient of
$\appTailBeta$ ($t = \appTailT$) against $\appTailBetaCalm$ in calm periods.

\emph{Stability of the regime dating.} If regimes are estimated, two checks are
worth their cost. Re-date the regimes from an entirely external source and
re-run; and, treating the regime path as a random object, draw paths from the
smoothing posterior of the fitted model and recompute. In the application the
central channel retains a statistic between $\appOfrLo$ and $\appOfrHi$ under
four external datings, and exceeds the conventional threshold in $\appFfbs$ per
cent of two hundred posterior paths.

\subsection{A reporting checklist}

The design layer can be summarised as six questions to answer before a
directional claim is made.
\begin{enumerate}\itemsep2pt
\item Is there a mechanical ordering of the observations, from clocks,
reporting lags, or publication schedules, that would produce this arrow with no
transmission?
\item Is there an observable at which the mechanical and the economic readings
disagree, and has the reading of each outcome been fixed in advance?
\item Does the observable the design hinges on survive an audit of the raw
data?
\item Does a timing placebo, in which the sender's information is already
public, return nothing?
\item Does the entire procedure, run on a reversible process with the same
nuisance features, reject at the nominal rate?
\item If regimes or windows are estimated, do the conclusions survive external
re-dating and posterior uncertainty in the dating?
\end{enumerate}

\section{Comparison with Existing Methods}\label{sec:comparison}

The methods this paper is most often compared with were built to answer
different questions, and much of the confusion in applied work comes from
reading an answer to one question as an answer to another. This section sets
out what each method actually estimates, scores them against ours on four
systems whose answers are known by construction, and states plainly where our
method is the weaker one.

\subsection{What each method estimates}

Table~\ref{tab:taxonomy} classifies the common tools by the feature of the
joint law they read. Definitions of each are collected in
Appendix~\ref{app:laboratory}, so that no external reference is needed.

\begin{table}[t]
\centering
\caption{What the common tools estimate.}
\label{tab:taxonomy}
\small
\begin{tabular}{>{\raggedright\arraybackslash}p{0.30\textwidth}>{\raggedright\arraybackslash}p{0.31\textwidth}>{\raggedright\arraybackslash}p{0.30\textwidth}}
\toprule
Method & Feature of the law it reads & Directional content \\
\midrule
Correlation, covariance networks, Mantegna distance, spanning trees,
principal components
& contemporaneous law only
& none, by Theorem~\ref{thm:ident}(i) \\[2pt]
Dynamic conditional correlation, rolling correlation, correlation-shift
contagion tests
& time variation in the contemporaneous law
& none \\[2pt]
Granger causality, transfer entropy (equivalent for Gaussian data)
& conditional predictability of $i$ from the past of $j$
& gross and confounded with persistence \\[2pt]
Forecast error variance connectedness, and its frequency decomposition
& shares of forecast error variance across an identification scheme
& gross, and not zero under reversibility \\[2pt]
Structural vector autoregressions, local projections
& impulse responses under an identifying restriction
& structural, conditional on the restriction \\[2pt]
Quantile and tail spillover measures
& conditional tail quantiles
& gross, tail-specific \\[2pt]
Convergent cross mapping, causal discovery on time series
& conditional independence or attractor reconstruction
& structural, under strong assumptions \\[2pt]
Circulation and its entropy-production component (this paper)
& antisymmetric part of the lagged law
& net, free of persistence confounding \\
\bottomrule
\end{tabular}
\begin{minipage}{0.92\textwidth}
\vspace{1ex}\footnotesize
Notes. Original sources, by row: correlation networks and the spanning tree
\citep{mantegna1999hierarchical}; dynamic conditional correlation
\citep{engle2002dynamic} and correlation-shift contagion tests
\citep{forbes2002contagion}; predictive causality \citep{granger1969investigating}
and transfer entropy \citep{schreiber2000measuring}, which coincide for
Gaussian data \citep{barnett2009granger}; forecast error variance connectedness
\citep{diebold2012better} built on the ordering-free decomposition of
\citet{pesaran1998generalized}, with its frequency version in
\citet{barunik2018measuring}; structural autoregressions
\citep{sims1980macroeconomics} and local projections \citep{jorda2005estimation};
convergent cross mapping \citep{sugihara2012detecting} and causal discovery
on time series \citep{runge2019inferring}. The decomposition of linear
dependence into directional feedbacks is due to \citet{geweke1982measurement}.
Definitions sufficient to reproduce each row are in
Appendix~\ref{app:laboratory}.
\end{minipage}
\end{table}

Two distinctions in that table do most of the work.

\emph{Net against gross.} The circulation is a net measure. If $i$ sends to $j$
and $j$ sends back with equal strength, the two cancel and $Q$ records nothing,
correctly, because the system is then reversible on that pair. Granger
causality, transfer entropy and connectedness are gross measures: they record
both flows and report both. Neither convention is better; they answer different
questions. An analyst asking how much of $j$'s variance originates elsewhere
wants a gross measure. An analyst asking which way the system turns wants a net
one. Reporting a gross measure and calling the larger of the two directions
``the direction'' is where the two get confused.

\emph{Confounded against confound-free.} As Remark~\ref{rem:te} showed, a
difference in own persistence produces a difference in predictability with no
difference in direction. Every gross predictive measure inherits this. The
circulation does not, because it compares
$\E[x_{i,t+\Delta}x_{j,t}]$ with $\E[x_{j,t+\Delta}x_{i,t}]$, two quantities
that reversibility equates exactly.

A third property is worth stating separately, because it is unusual.

\begin{proposition}[invariance]\label{prop:invariance}
The magnitude $\sigma_Q$ is invariant under every invertible linear
reparameterisation of the series: if $\tilde x_t = M x_t$ with $M$ invertible,
then $\tilde\sigma_Q = \sigma_Q$.
\end{proposition}

The proof is one line and is given in Appendix~\ref{app:proofs}; numerically,
$\sigma_Q$ changes by at most $10^{\invMaxRelOrder}$ in relative terms across
$\invDraws$ random invertible reparameterisations. The practical content is
that the magnitude does not depend on the units of the series, on whether one
studies levels or a rotation of them, or on which basis of the same
$n$-dimensional space the analyst happens to use. Correlation-based measures
are invariant only to rescaling of individual series, and forecast error
variance decompositions are not invariant at all: they depend on the
identification scheme, which is why the ordering-free generalised version was
introduced and why its shares do not sum to one before normalisation.

\subsection{Four test systems with known answers}

Comparing methods on real data cannot settle anything, because the truth is
unavailable. We therefore use four systems in which it is known by
construction. Full definitions are in Appendix~\ref{app:laboratory}; the
constructions are all elementary.

\begin{description}\itemsep3pt
\item[A, reversible.] A stationary linear system whose drift is purely
gradient, so $Q = 0$ exactly and no directional claim of any kind is warranted.
The series are correlated, and their persistences differ.
\item[B, circulating.] The same stationary law as A, with a circulating
component added to the dynamics. By construction A and B have identical
covariance matrices, so every contemporaneous method must return the same
answer for both.
\item[C, staggered clocks.] Three markets recording a single common news stream
at three different hours. There is no transmission whatsoever, and there is a
real arrow of time.
\item[D, nonlinear.] $y_t = c(x_{t-1}^2 - 1) + e_t$ with $x$ independent
draws: transmission from $x$ to $y$ is total and one-directional, and every
lagged covariance is exactly zero.
\end{description}

\begin{figure}[t]
\centering
\includegraphics[width=0.92\textwidth]{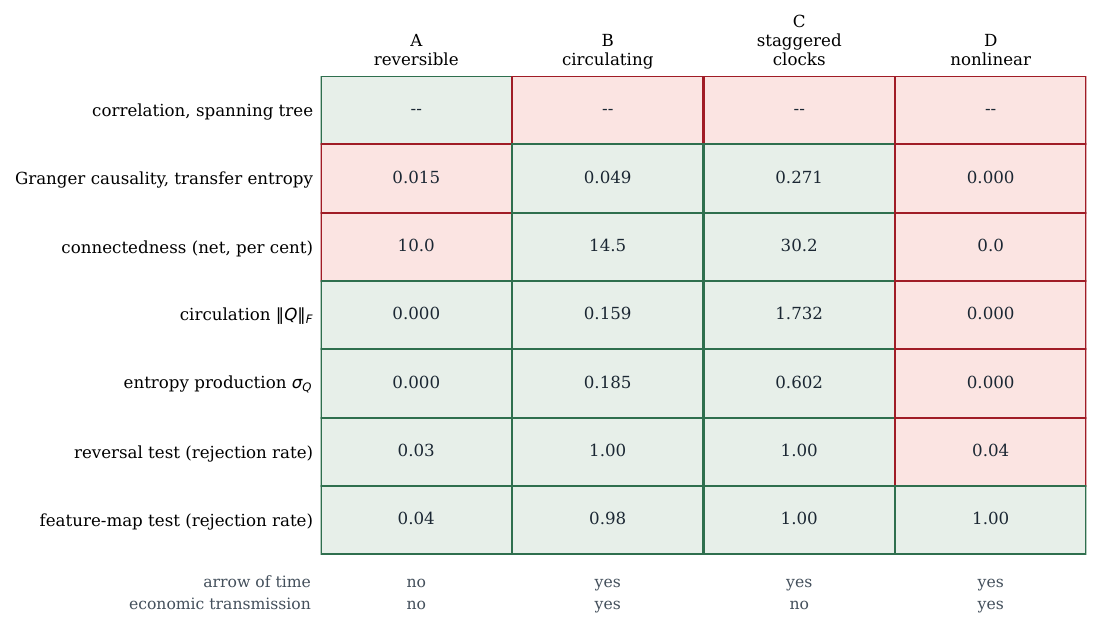}
\caption{The same comparison, with each cell shaded according to whether the
method's reading, taken as a statement about the arrow of time, matches the
truth. Green marks agreement, red disagreement. The population rows are
descriptors rather than tests, and a red cell means the reading misleads when
taken as a direction, not that the method errs on its own question. No method
is correct in every column, and the last two columns are correct for different
reasons.}
\label{fig:scoreboard}
\end{figure}

Table~\ref{tab:scoreboard} and Figure~\ref{fig:scoreboard} report what each
method says. Population values are computed exactly from the systems'
definitions; the test rows are rejection rates at the five per cent level over
$\labM$ samples of $\labT$ observations.

\begin{table}[t]
\centering
\caption{What each method reports on four systems with known answers.}
\label{tab:scoreboard}
\small
\begin{tabular}{lcccc}
\toprule
 & A & B & C & D \\
 & reversible & circulating & staggered & nonlinear \\
\midrule
Truth: is there an arrow of time? & no & yes & yes & yes \\
Truth: is there transmission? & no & yes & no & yes \\
\midrule
Correlation, distance, spanning tree & -- & -- & -- & -- \\
Transfer entropy asymmetry (nats) & 0.015 & 0.049 & 0.271 & 0.000 \\
Connectedness, net (per cent) & 10.0 & 14.5 & 30.2 & 0.0 \\
Circulation $\lVert Q_\Delta\rVert_F$ & 0.000 & 0.159 & 1.732 & 0.000 \\
Entropy-production component $\sigma_Q$ (nats) & 0.000 & 0.185 & 0.602 & 0.000 \\
\midrule
Granger test, any pair rejected & 1.00 & 1.00 & 1.00 & 0.13 \\
Reversal test, linear statistic & 0.03 & 1.00 & 1.00 & 0.04 \\
Reversal test, feature statistic & 0.04 & 0.98 & 1.00 & 1.00 \\
\bottomrule
\end{tabular}

\begin{minipage}{0.92\textwidth}
\vspace{1ex}\footnotesize
Notes. Rows three to seven are population values, computed from the systems'
exact moments; rows eight to ten are rejection frequencies at the five per cent
level over $\labM$ replications of length $\labT$, with $\labR$ randomisation
draws each. The Granger row reports the frequency with which any ordered pair
is declared significant after a Bonferroni correction across the six pairs.
Connectedness is the largest absolute net share in per cent.
\end{minipage}
\end{table}

\emph{System A.} There is no direction here at all, and
the correlation-based methods are correctly silent, since they say nothing
about direction in any system. Granger causality rejects for at least one
ordered pair in $\labGrangerA$ of samples even after a Bonferroni correction.
The population transfer entropy asymmetry is $\labTeA$ nats, and the
connectedness index reports a largest net directional share of $\labDyA$ per
cent, naming the second series as the dominant net transmitter; a nonzero net
share under reversibility is a mechanical property of a variance-share
accounting when shocks are correlated and persistences differ, not a failure
of the decomposition on its own terms. Circulation and
entropy production are zero to machine precision, $10^{\labSigAOrder}$, and the
reversal test rejects in $\labRevA$ of samples against a nominal five per cent.
The lesson is not that Granger causality is broken; it is estimating what it
claims to estimate, namely a difference in predictability. The lesson is that a
difference in predictability is not a direction.

\emph{System B.} B was built to
have exactly the same stationary law as A. The two correlation matrices agree
to $10^{\labCorrGapOrder}$ and their minimum spanning trees are
\labMstIdentical. Any statistic computed from the contemporaneous law therefore
gives the identical answer in a system with no direction and in a system that
circulates. Meanwhile the circulation is $\labQfB$, the magnitude is $\labSigB$
nats, and the reversal test rejects in $\labRevB$ of samples. Connectedness
also reports a net flow here, $\labDyB$ per cent, but it named a different
series as the dominant transmitter than it did in A, which is the point: it
produces a directional ranking in both, with no accompanying test, and the
ranking is not informative about which of the two systems has a direction.

\emph{System C.} All the directional methods fire,
including ours, and all of them are right that the sampled law has an arrow.
None of them can tell that no transmission occurred, and none of them should be
expected to. This is the case for the design layer of
Section~\ref{sec:design}.

\emph{System D.} Here every lagged covariance is exactly
zero, so the circulation, the magnitude, the transfer entropy asymmetry and the
connectedness net flow are all exactly zero, and the linear reversal test
rejects in $\labRevD$ of samples, which is the nominal rate. Transmission is
nonetheless complete and one-directional. The linear Granger test picks it up in
$\labGrangerD$ of samples, better than nothing but far from reliable, since the
dependence is invisible to a linear projection. The feature-map statistic of
Section~\ref{sec:framework} rejects in $\labFeatD$ of samples while remaining
correctly sized in system A, where it rejects in $\labFeatA$.

\subsection{Limitations relative to the alternatives}

The following limitations are structural, not matters of implementation.

\emph{Net measurement.} The circulation is a net measure, so flows that
cancel are invisible to it. If the intended
question is how much of one series' variance originates in another, regardless
of direction, a variance decomposition is the right tool and the circulation is
not.

\emph{Second moments.} The linear reading sees only second moments unless the
analyst extends it, and System D is the
warning. The feature-map extension is available and costs nothing in validity,
but it multiplies the number of parameters, and with a rich basis the bias
correction of Section~\ref{sec:estimation} becomes essential rather than
advisable.

\emph{Stationarity.} The method requires stationarity within the estimation
window. A system whose
parameters drift will show circulation that reflects the drift. The remedy is
to define windows or regimes externally and to check the sensitivity of the
conclusions to that definition; the remedy is not to shorten the window until
the drift is invisible, since $n^2/T$ then becomes the binding problem.

\emph{Sampling.} The method cannot see faster than the sampling interval, and
there is a window of visibility; Section~\ref{sec:sampling} quantifies both.

\emph{Reduced form.} The circulation does not identify structure; it is a
reduced-form object.
It does not deliver impulse responses, counterfactuals, or policy experiments,
and mapping it to a structural model requires exactly the identifying
assumptions that structural work always requires. What it does is tell the
analyst whether there is any directional structure to identify.

\emph{Within-interval transmission.} The method says nothing about
transmission inside the sampling interval. Contemporaneous transmission, completed before the next observation, appears in
the innovation covariance $\Sigma_\varepsilon$ and carries no direction at
this frequency. Two markets that respond to each other within the hour, on
daily data, are simply correlated.

\emph{Dimension.} The dimension enters as $n^2/T$: at twenty-six series and
eight hundred
observations the correction is essential; at one hundred series and the same
length no estimator in this paper is trustworthy without regularisation, which
we have not developed.

\subsection{Complementarities}

None of this argues for replacing the standard toolkit, and the useful reading
is that the tools answer complementary questions. A practical division of
labour: use correlation networks and their dynamic extensions to describe
grouping and co-movement, which is what they measure; use variance
decompositions to size the total exposure of one series to the rest; use
structural methods when an identifying restriction is genuinely available; and
use the circulation to answer whether there is a direction at all, how large it
is in units that do not depend on the parameterisation, and where it sits in
the system. When the circulation is zero, a directional reading of any of the
other measures should be treated as a statement about predictability or about
variance shares, not about direction.

\section{Implementation}\label{sec:implementation}

\subsection{Two reference implementations}

The method is deliberately small: the whole estimator is a handful of matrix
operations, and the whole test is a loop that reverses stretches of a matrix
and recomputes. There are two reference implementations, written separately
so that the estimator core of each could be run against the other.

The first is a Python module depending only on \texttt{numpy} and
\texttt{scipy}. The second is an R package, \texttt{arrowoftime}, written in
base R with \texttt{stats} as its only import, laid out in the standard
package structure with documented functions, examples, unit tests, two
vignettes, and the data of the application; it passes \texttt{R CMD check
-{}-as-cran} with no errors and no warnings. Table~\ref{tab:api} lists the
entry points. The R package implements the full interface; the Python module
implements the estimator and inference core under closely corresponding names,
and the two are compared entry by entry on the same data below.

\begin{table}[t]
\centering
\caption{The interface of the R package. The Python module covers the
estimator and inference core (moments, circulation, cross-fitting, reversal
test, jackknife, stepdown, simulation); the session decomposition and the
data-preparation helpers are R only.}
\label{tab:api}
\small
\begin{tabular}{>{\raggedright\arraybackslash}p{0.30\textwidth}>{\raggedright\arraybackslash}p{0.62\textwidth}}
\toprule
Function & Returns \\
\midrule
\texttt{aot\_moments} & the two moment matrices $\widehat S$ and
$\widehat C(\Delta)$, pooled within episodes \\
\texttt{aot\_circulation} & $\widehat Q_\Delta$, its norm, net flows, the
implied $\widehat\Phi$ and $\widehat\Sigma_\varepsilon$, the plug-in magnitude \\
\texttt{aot\_sigma\_crossfit} & the cross-fitted magnitude \eqref{eq:cf} \\
\texttt{aot\_reversal\_test} & observed statistics, joint null draws, and
$p$-values for the regime-and-scale grid of cells \\
\texttt{aot\_jackknife} & delete-one-block standard error and interval \\
\texttt{aot\_romano\_wolf} & stepdown adjusted $p$-values from joint draws \\
\texttt{aot\_shift\_test} & the conditional circular-shift test of
Algorithm~\ref{alg:shift} \\
\texttt{aot\_ladder\_frame}, \texttt{aot\_ladder\_channels} & the session
decomposition of Section~\ref{sec:design} \\
\texttt{aot\_episodes}, \texttt{aot\_winsorise}, \texttt{aot\_modwt\_details} &
regime episodes, winsorisation, scale decomposition \\
\texttt{aot\_simulate\_var1} & a stationary system with a chosen circulation \\
\bottomrule
\end{tabular}
\end{table}

\subsection{A complete example}

The following is the entire analysis of a panel: the circulation, the net
flows, the corrected magnitude, and an exact test.

\begin{lstlisting}[language=Python]
import numpy as np
from aotlib import circulation_stats, sigma_crossfit, reversal_test

X = ...                                  # T by n array, one row per period
Z = (X - X.mean(0)) / X.std(0, ddof=1)   # scale within the estimation window

st  = circulation_stats([Z])             # Q, net flows, plug-in magnitude
sig = sigma_crossfit([Z], 50)            # the bias-corrected magnitude

def stat(W):                             # any statistic may be used
    Y = (W - W.mean(0)) / W.std(0, ddof=1)
    return np.linalg.norm(circulation_stats([Y])["Q"], "fro")

res = reversal_test(X, stat, R=999, block_length=100, seed=1)
print(st["net"], sig, res["p"])
\end{lstlisting}

\noindent The same analysis in R:

\begin{lstlisting}[language=R]
library(arrowoftime)

Z  <- scale(X)                           # X is a T by n matrix
st <- aot_circulation(list(Z))           # Q, net flows, plug-in magnitude
cf <- aot_sigma_crossfit(list(Z), fold_length = 50)

st$Q ; st$net ; c(plugin = st$sigma_plugin, crossfit = cf)
\end{lstlisting}

Run on $\exT$ observations of a three-market system that circulates from the
first market to the second, to the third, and back, the listing returns a
circulation norm of $\exQfrob$, net flows
$(\exNetOne, \exNetTwo, \exNetThree)$, a plug-in magnitude of $\exPlugin$
against a corrected magnitude of $\exCf$ nats per period, and an exact
$p$-value of $\exP$ against the reversibility null. On the same input the two
implementations agree to $10^{\exGapOrder}$ across every reported quantity,
which is the precision at which the comparison was made.

\subsection{A worked example}

Table~\ref{tab:worked} gives every intermediate quantity for a system of three
markets, labelled East, Centre and West, simulated for $\weT$ periods from a
known circulation that runs East to Centre to West and back.

\begin{table}[t]
\centering
\caption{The worked example: estimated moments, estimated circulation, and the
circulation that generated the data.}
\label{tab:worked}
\small
\begin{tabular}{lccc|ccc}
\toprule
 & \multicolumn{3}{c|}{$\widehat{S}$} & \multicolumn{3}{c}{$\widehat{C}(\Delta)$} \\
 & East & Centre & West & East & Centre & West \\
\midrule
East & +0.999 & +0.499 & +0.303 & +0.636 & +0.378 & +0.361 \\
Centre & +0.499 & +0.999 & +0.363 & +0.558 & +0.727 & +0.286 \\
West & +0.303 & +0.363 & +0.999 & +0.253 & +0.422 & +0.530 \\
\midrule
 & \multicolumn{3}{c|}{$\widehat{Q}_\Delta$} & \multicolumn{3}{c}{$Q_\Delta$ (truth)} \\
East & +0.000 & -0.090 & +0.054 & +0.000 & -0.077 & +0.046 \\
Centre & +0.090 & +0.000 & -0.068 & +0.077 & +0.000 & -0.067 \\
West & -0.054 & +0.068 & +0.000 & -0.046 & +0.067 & +0.000 \\
\bottomrule
\end{tabular}

\begin{minipage}{0.94\textwidth}
\vspace{1ex}\footnotesize
Notes. $\widehat S$ is the contemporaneous covariance of the standardised
series and $\widehat C(\Delta)$ the lag-one covariance;
$\widehat Q_\Delta$ is the antisymmetric part of the latter. A positive entry
in row $i$ and column $j$ means $j$ leads $i$.
\end{minipage}
\end{table}

Every step can be checked with a calculator. The lag-one covariance of East
today with Centre yesterday is $\weCtwelve$; of Centre today with East
yesterday, $\weCtwentyone$; half their difference is $\weQtwelve$, which is the
entry of $\widehat Q_\Delta$ in the first row and second column. Summing the
first row and doubling gives the net flow of East, $\weNetEast$, against a
true value of $\weNetTrueEast$; the three net flows sum to zero by
construction.

The magnitudes illustrate Section~\ref{sec:estimation} at a scale where the
correction is small and can therefore be seen clearly. The truth is $\weSigTrue$
nats per period. The plug-in returns $\wePlugin$ and the cross-fitted estimator
$\weCf$, a gap of $\weBiasGap$, against the predicted bias $n(n-1)/T =
\weBiasPred$. The delete-block jackknife, with $150$-observation blocks here, gives a
standard error of $\weJkSe$ and an interval $[\weJkLo, \weJkHi]$, which
contains both the estimate and the truth; in this single draw the cross-fitted
estimate sits $\weCfDevSe$ standard errors above the truth, which is what a
single draw is entitled to do. The exact reversal test over $\weR$ draws returns $p = \weP$. With three
series and $\weT$ observations the plug-in bias is a rounding error; with
twenty-six series and eight hundred observations, as in
Section~\ref{sec:application}, the same quantity is larger than the estimate.

\subsection{Computational cost}

The estimator is $O(T n^2)$ per evaluation and the randomisation test repeats
it $R$ times, so the whole procedure is $O(R\,T\,n^2)$ with no matrix inversion
inside the loop beyond an $n \times n$ solve. On one core of an ordinary
workstation, a panel of twenty-six series and eight hundred observations with
$R = 1{,}000$ draws completes in well under a minute, and the whole of the
application of Section~\ref{sec:application}, including four time scales, two
regimes, jackknife standard errors, and stepdown adjustment, runs in under
fifteen minutes. The expensive components in practice are not the estimator but
the studies that surround it: an end-to-end placebo that repeats the entire
procedure a hundred times, and nonparametric diagnostics with nearest-neighbour
estimators.

Two implementation notes save time. First, the statistic inside the
randomisation loop should be written to recompute only what depends on the
data, but everything that does depend on the data must be recomputed, including
filtering and standardisation. Second, when several statistics are tested, they
should share the draws: this is both cheaper and necessary for the stepdown
adjustment of Section~\ref{sec:inference}.

\subsection{Materials}

The materials accompanying this paper contain the Python module, the numbered
scripts that produce the simulation and laboratory results and every figure,
the R package as a source archive, and the datasets of the application in
plain text; the application's cell-level results are imported from the
companion study's deposited output and reduced, not recomputed, by these
scripts. Each script writes
its results to a structured file, and the manuscript reads its numbers from
those files, so that every estimate, test statistic, and $p$-value printed in
the text is read from a results file rather than typed. Random seeds,
draw counts, fold lengths, and block lengths are recorded alongside the results
they produced.

\subsection{A recommended workflow}\label{sec:workflow}

The steps below assemble the paper into a procedure. Each points to the
section that justifies it.

\begin{enumerate}\itemsep2pt
\item Fix the estimation windows or regimes by an external criterion and
check stationarity within them (Sections~\ref{sec:estimation}
and~\ref{sec:design}).
\item Check that the sampling interval sits inside the window of visibility:
fit the transition matrix, read off the implied half-lives, and compare them
with the observation interval (Section~\ref{sec:sampling}).
\item Estimate $S$ and $C(\Delta)$ within episodes, with no lagged product
crossing an episode boundary (Section~\ref{sec:estimation}).
\item Read the circulation $\widehat Q_\Delta$ and the net flows directly;
they are unbiased (Lemma~\ref{lem:bias}).
\item Estimate the magnitude by cross-fitting, never by the plug-in, and
attach block-jackknife standard errors (Section~\ref{sec:estimation}).
\item Test against the reversal null, applying the reversals to the rawest
available record and recomputing every non-invariant step inside the loop
(Section~\ref{sec:inference}).
\item If several cells are tested, adjust with the stepdown procedure on the
same draws (Section~\ref{sec:inference}).
\item If nonlinear transmission is plausible, repeat with feature maps; the
same draws remain valid (Sections~\ref{sec:framework}
and~\ref{sec:inference}).
\item Before interpreting a detected arrow as transmission, rule out
mechanical orderings by design, not by statistics
(Section~\ref{sec:design}).
\item Only then, and only with additional identifying assumptions, formulate
structural claims; the circulation itself does not supply them
(Section~\ref{sec:comparison}).
\end{enumerate}

A null result needs the same care as a detection: by
Section~\ref{sec:sampling}, an absent arrow is evidence of reversibility only
when step 2 has confirmed that an arrow of the relevant speed would have been
visible at the sampling interval used.

\section{Empirical Application}\label{sec:application}

This section summarises an application, both to show the method operating end
to end and because the results are the reason we built it. The full study, with
its data appendix and every robustness layer, is a companion paper
\citep{bhandari2026arrow}; here we report only what illustrates the method.

\subsection{Data and setting}

The panel is $\appN$ asset classes, fourteen equity indices, five commodities
and seven currencies against the dollar, in daily log returns from January 2007
to November 2025. Volatility regimes are identified from the data by a
two-state hidden Markov model fitted to a global stress factor, and crisis
episodes are collected to a fixed budget of eight hundred days, with the single
longest calm run truncated to the same length so that the two regime panels are
matched by construction. Each series is decomposed into four time scales by a
wavelet transform, and the estimator of Section~\ref{sec:estimation} is applied
to every regime and scale cell, with reversal tests applied to the raw returns
before filtering and delete-block jackknife standard errors. Following a rule
committed before estimation in the companion study, the three finest scales
are inferential and the coarsest, whose effective sample is an order of
magnitude smaller, is reported as descriptive only; the stepdown adjustment
runs across the six inferential cells. The regime dating is estimated once on
the raw panel and held fixed across the reversal draws, so the cell-level
tests are conditional on the dating; the placebo described below is the check
that this conditioning is innocuous.

\subsection{Results}

\begin{figure}[t]
\centering
\includegraphics[width=\textwidth]{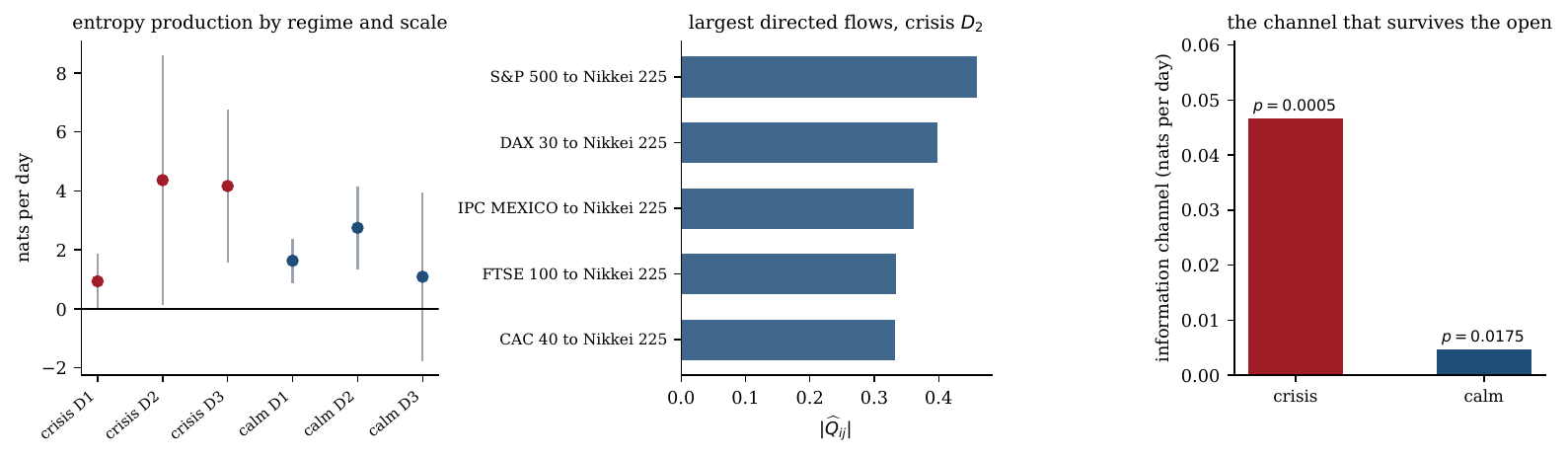}
\caption{The application. Left: the cross-fitted entropy-production component
$\widehat\sigma_Q$ by regime and scale with jackknife intervals. Centre: the five largest directed flows at the
dominant scale in the crisis regime. Right: the channel that survives the
receiving market's opening auction, by regime, with the $p$-values of the
conditional circular-shift test.}
\label{fig:application}
\end{figure}

Table~\ref{tab:application} reports the estimates cell by cell and
Figure~\ref{fig:application} shows them alongside the leading flows and the
session split. Three results illustrate three different parts of the method.

\begin{table}[t]
\centering
\caption{Circulation and the quadratic entropy-production component $\sigma_Q$ by
regime and time scale.}
\label{tab:application}
\small
\begin{tabular}{lrrrrrrr}
\toprule
cell & $T$ & $\lVert\widehat{Q}_\Delta\rVert_F$ & s.e. & $p_{\mathrm{rev}}$ & $p_{\mathrm{RW}}$ & $\widehat{\sigma}_Q$ & s.e. \\
\midrule
crisis $D_1$ & 799 & 1.90 & 0.43 & 0.007 & 0.008 & 0.93 & 0.45 \\
crisis $D_2$ & 799 & 3.16 & 1.06 & 0.007 & 0.008 & 4.36 & 2.14 \\
crisis $D_3$ & 799 & 1.90 & 0.62 & 0.062 & 0.062 & 4.16 & 1.31 \\
crisis $D_4$ & 799 & 1.22 & 0.43 & 0.177 & -- & 15.74 & 10.84 \\
calm $D_1$ & 800 & 1.72 & 0.30 & 0.008 & 0.008 & 1.63 & 0.36 \\
calm $D_2$ & 800 & 2.14 & 0.41 & 0.023 & 0.023 & 2.75 & 0.70 \\
calm $D_3$ & 800 & 1.29 & 0.21 & 0.004 & 0.050 & 1.09 & 1.44 \\
calm $D_4$ & 800 & 1.00 & 0.14 & 0.006 & -- & 10.11 & 4.68 \\
\bottomrule
\end{tabular}

\begin{minipage}{0.86\textwidth}
\vspace{1ex}\footnotesize
Notes. $T$ is the number of regime days. Standard errors are delete-one-block
jackknife with hundred-day blocks within episodes. $p_{\mathrm{rev}}$ is the
block time-reversal randomisation test applied to raw returns, unadjusted;
$p_{\mathrm{RW}}$ is the stepdown familywise adjustment across the six
inferential cells, with the two descriptive $D_4$ cells outside the family.
$\widehat\sigma_Q$ is cross-fitted and reported in nats per day.
\end{minipage}
\end{table}

The first result is that the arrow is permanent, which the design was not
built to expect. After the familywise adjustment, the reversibility null is
rejected at the two finest crisis scales ($p_{\mathrm{RW}} = \appPAdjCrOne$
and $\appPAdjCrTwo$) and, contrary to the natural prior that direction is a
crisis phenomenon, in calm as well: $p_{\mathrm{RW}} = \appPAdjCaOne$ and
$\appPAdjCaTwo$ at the two finest calm scales, with the third calm scale
sitting exactly at the five per cent line ($p_{\mathrm{RW}} = \appPAdjCaThree$
from an unadjusted $\appPCaThree$). The descriptive coarse cells point the
same way ($p = \appPCaFour$ unadjusted in calm) but are outside the registered
family and carry no inferential weight. The estimated circulation has a clear
structure:
its largest entries run from developed Western equity markets into the Tokyo
session, in both regimes, with the single largest flow from the S\&P~500 into
the Nikkei~225. Direction in global markets is organised by the trading clock.

The second result is that the magnitude needed the correction. At the dominant
scale in crisis the cross-fitted magnitude is $\appSigCrTwo$ nats per day,
equivalently $\appSigPerPairCrTwo$ millinats per pair across the
$\appNPairs$ pairs of the panel. The plug-in estimator on the same cell
returns $\appPluginCrTwo$, a factor of $\appPluginRatioCrTwo$ larger. The gap
exceeds the unit-weight rule of thumb $n^2/T = \gridRatioBig$ severalfold
because the weighted bias constant of Lemma~\ref{lem:bias} depends on the
conditioning of the innovation covariance, which deteriorates at filtered
scales; the like-for-like calibration in Appendix~\ref{app:estimation}, built
to match this cell's own moments, reproduces exactly this behaviour. An
uncorrected comparison of magnitudes across regimes of unequal length would
be reporting sample sizes.

The third result is that the design separated mechanics from transmission.
Applying
Section~\ref{sec:design} to the receiving session, the United States channel
into Tokyo turns out to be almost entirely price discovery at the open: the
settlement component is enormous in both regimes ($\appSettleTe$ nats,
$t = \appSettleT$ in crisis) while the information component is statistically
zero ($t = \appSettleInfoT$). The exception is European. In crisis the
information channel from the European session into the following Tokyo session
is $\appTeCrisis$ nats with $t = \appTCrisis$ and conditional randomisation
$p = \appPShiftCrisis$, against $\appTeCalm$ nats
($t = \appTCalm$, $p = \appPShiftCalm$) in calm: a ratio of $\appRatio$. The
channel is carried by the largest decile of European shocks, survives four
external re-datings of the crisis regime with a statistic between $\appOfrLo$
and $\appOfrHi$, and holds in $\appFfbs$ per cent of two hundred posterior
regime paths.

\subsection{Methodological lessons from the application}

Two features of the exercise are worth extracting, because they generalise.

The aggregate and the channel disagree, and the method makes the disagreement
visible rather than letting one absorb the other. The crisis and calm
magnitudes differ in the direction the underlying economics predicts, but with
jackknife uncertainty the aggregate difference is not significant at
conventional levels. What is significant and robust is the regime switch in one
specific channel and its concentration in one tail. Had only the aggregate been
reported, the finding would have been recorded as a failure; had only the
channel been reported, it would have been overstated as an economy-wide
phenomenon.

The end-to-end placebo was worth its cost. Passing a reversible process with
fitted volatility dynamics through the entire procedure, including the regime
estimation, gives rejection rates between $\appPlcMin$ and $\appPlcMax$ against
a nominal five per cent over $\appPlcM$ replications. Since the regime dating is
itself estimated from the data, no component test could have established this.

\section{Scope and Limitations}\label{sec:limitations}

We have stated the limitations where they arose. This section collects them in
one place, adds the ones that are matters of scope rather than of method, and
lists what we regard as the most useful open problems.

\subsection{Limitations of the estimand}

The structural limitations of the estimand itself, that it is a net measure
blind to balanced flows, that its linear form reads second moments only, that
nothing faster than the sampling interval is visible, that it is a
reduced-form object identifying no mechanism, and that a steady state is
assumed within the estimation window so that parameter drift will read as
circulation, were each developed where they arose and are collected in
Section~\ref{sec:comparison}. One deserves restating because it governs
research design: regime and window definitions carry real weight, and should
be external, or estimated and then stress-tested as in
Section~\ref{sec:design}.

\subsection{Limitations of the inference}

The reversal test is exact under its null, which is reversibility of the
within-block law together with independence across blocks. Both parts can fail.
Dependence that survives beyond the block length degrades the test at rate
$O(m/L)$, so a block length that is short relative to the memory of the series
is a real error, not a conservative choice. And the reversal test is a test of
reversibility, not of any particular alternative: rejection says the record and
its reverse differ, and the reason may be cross-series transmission, a single
series with an asymmetric volatility response, or a filter that was not
recomputed inside the loop.

The conditional circular-shift test of Section~\ref{sec:inference} is not
exact even under its null: the rotation joins the end of the record to its
beginning, and the resulting wrap-around error vanishes only with the sample
length. At the application's sample sizes the error is negligible, but the
test should not be described as exact, and we have not done so.

The jackknife intervals under-cover slightly in the calibration of
Appendix~\ref{app:estimation}, at $\covJk$ against a nominal $0.95$, because
episodes provide few blocks. Where coverage matters more than convenience,
the exact randomisation test should carry the inferential weight and the
interval should be read as a descriptive range.

\subsection{Scope conditions}

The method is built for a panel of comparable series observed on a common grid,
in a steady state, with the number of series small relative to the number of
observations. It is not built for, and should not be used for, event studies
with a handful of observations, panels with hundreds of series and short
windows, irregularly sampled or asynchronous records without an explicit
alignment step, or series with unit roots. Series should be differenced or
otherwise rendered stationary before anything in this paper is applied; the
circulation of a set of random walks is not a meaningful object.

\subsection{Open problems}

\emph{High dimension.} At $n^2/T$ above one, no estimator here is reliable.
Regularised versions of the quadratic functional, along the lines developed for
covariance estimation, would extend the reach considerably, and the
cross-fitting structure is already the right one to build on.

\emph{Choosing the basis.} Proposition~\ref{prop:feature} says that a complete
basis characterises reversibility, and says nothing about which finite basis to
use with a given sample. A data-driven choice with a valid post-selection
reference distribution would be genuinely useful, and the randomisation
framework makes it feasible in principle, since the selection can be repeated
inside the loop.

\emph{Time-varying circulation.} We estimate a constant circulation within
regimes. A smoothly varying $Q_\Delta(t)$, estimated by local moments with an
inferential layer that respects the smoothing, is the natural next object.

\emph{Continuous-time and mixed-frequency records.} Assets observed at
different frequencies, or asynchronously, are the norm in practice, and the
alignment step we use, the session decomposition of Section~\ref{sec:design}, is
specific to markets with defined trading hours. A general treatment of
asynchronous observation within this framework is open.

\emph{Interventions.} The strongest version of a directional claim is
counterfactual, and the circulation is not counterfactual. What conditions turn
a measured circulation into an intervention-relevant object is, in our view, the
most important open question in this line of work, and it is one that
identification-first thinking is well placed to state precisely even where it
cannot yet answer it.

\section{Conclusion}\label{sec:conclusion}

The question of which series leads which is asked constantly and answered
loosely. This paper has argued that it has a precise answer, that the answer is
the same in every field where the question arises, and that it is not the answer
the standard toolkit provides.

The precise answer is that direction, for a system in a steady state, is time
irreversibility, and that the whole of its second-order trace is carried by
the antisymmetric part of the lagged covariance, with feature maps extending
the same construction to whatever lies beyond second moments. That object has three properties which together make
it usable. It is invisible to every method built on contemporaneous
co-movement, which is why the correlation networks that dominate applied work
cannot answer the question no matter how much data they are given. It is free
of the confounding between direction and persistence that afflicts every
measure built on predictive asymmetry, which is why two markets can differ in
how well each predicts the other while the system as a whole has no direction
at all. And its natural magnitude, under the Gaussian benchmark, is an
entropy-production functional of the identified circulation, the quadratic
component of the divergence rate between the forward record and the reversed
one, measured in nats per period, invariant to how the series are
parameterised, and interpretable without reference to the units of the data.

Around that estimand the paper has built what applied work needs: an estimator
whose first-order bias is removed rather than assumed away, standard errors
that respect the structure that makes the correction work, a test whose
validity comes from the symmetry in the null rather than from an asymptotic
approximation, a multiplicity correction that costs nothing, an extension to
nonlinear structure that leaves the exactness intact, a sampling theory that
tells the analyst in advance whether the exercise can succeed, and a design
layer that separates a measured arrow from an economic claim.

We have also been explicit about where the method fails. It is a net measure
and cannot see balanced flows. Its linear form is exactly blind to purely
nonlinear transmission, and we exhibited a system where it reports zero while
transmission is total. It cannot distinguish the rotation of the earth from the
propagation of news, and no observational method can. It says nothing about
what would happen under an intervention. A method that is clear about these
things is more useful than one that is not, and the comparison of
Section~\ref{sec:comparison} was constructed to make each failure visible,
including our own.

The application that motivated the work found that global markets carry a
permanent arrow of time, organised by the trading clock, that on ordinary days
the receiving market's opening auction closes it completely, and that in crisis
one channel stays open past the open and is carried by the largest foreign
shocks. That is a substantive finding, and it is not measurable inside the
correlation paradigm. It is also, we think, an example of a general pattern: the
questions applied researchers most want to answer about coupled systems are
directional, and directional questions have a precise home in the arrow of
time.

\bibliographystyle{plainnat}
\bibliography{refs}

\begin{thebibliography}{51}
\providecommand{\natexlab}[1]{#1}
\providecommand{\url}[1]{\texttt{#1}}
\expandafter\ifx\csname urlstyle\endcsname\relax
  \providecommand{\doi}[1]{doi: #1}\else
  \providecommand{\doi}{doi: \begingroup \urlstyle{rm}\Url}\fi

\bibitem[Barnett et~al.(2009)Barnett, Barrett, and Seth]{barnett2009granger}
Lionel Barnett, Adam~B. Barrett, and Anil~K. Seth.
\newblock Granger causality and transfer entropy are equivalent for {Gaussian}
  variables.
\newblock \emph{Physical Review Letters}, 103\penalty0 (23):\penalty0 238701,
  2009.
\newblock \doi{10.1103/PhysRevLett.103.238701}.

\bibitem[Barun{\'i}k and K{\v{r}}ehl{\'i}k(2018)]{barunik2018measuring}
Jozef Barun{\'i}k and Tom{\'a}{\v{s}} K{\v{r}}ehl{\'i}k.
\newblock Measuring the frequency dynamics of financial connectedness and
  systemic risk.
\newblock \emph{Journal of Financial Econometrics}, 16\penalty0 (2):\penalty0
  271--296, 2018.
\newblock \doi{10.1093/jjfinec/nby001}.

\bibitem[Battle et~al.(2016)Battle, Broedersz, Fakhri, Geyer, Howard, Schmidt,
  and MacKintosh]{battle2016broken}
Christopher Battle, Chase~P. Broedersz, Nikta Fakhri, Veikko~F. Geyer, Jonathon
  Howard, Christoph~F. Schmidt, and Fred~C. MacKintosh.
\newblock Broken detailed balance at mesoscopic scales in active biological
  systems.
\newblock \emph{Science}, 352\penalty0 (6285):\penalty0 604--607, 2016.
\newblock \doi{10.1126/science.aac8167}.

\bibitem[Beare and Seo(2014)]{beare2014time}
Brendan~K. Beare and Juwon Seo.
\newblock Time irreversible copula-based {Markov} models.
\newblock \emph{Econometric Theory}, 30\penalty0 (5):\penalty0 923--960, 2014.
\newblock \doi{10.1017/S0266466614000115}.

\bibitem[Bhandari et~al.(2026)Bhandari, Parida, and Sahu]{bhandari2026arrow}
Avishek Bhandari, Ipsita Parida, and Hitesh~Kumar Sahu.
\newblock The arrow of time in global financial markets: Identification,
  entropy production, and crisis transmission.
\newblock Working paper, Indian Institute of Technology Bhubaneswar, 2026.

\bibitem[Chen et~al.(2000)Chen, Chou, and Kuan]{chen2000testing}
Yi-Ting Chen, Ray~Y. Chou, and Chung-Ming Kuan.
\newblock Testing time reversibility without moment restrictions.
\newblock \emph{Journal of Econometrics}, 95\penalty0 (1):\penalty0 199--218,
  2000.
\newblock \doi{10.1016/S0304-4076(99)00036-6}.

\bibitem[Chernozhukov et~al.(2018)Chernozhukov, Chetverikov, Demirer, Duflo,
  Hansen, Newey, and Robins]{chernozhukov2018double}
Victor Chernozhukov, Denis Chetverikov, Mert Demirer, Esther Duflo, Christian
  Hansen, Whitney Newey, and James Robins.
\newblock Double/debiased machine learning for treatment and structural
  parameters.
\newblock \emph{The Econometrics Journal}, 21\penalty0 (1):\penalty0 C1--C68,
  2018.
\newblock \doi{10.1111/ectj.12097}.

\bibitem[Darolles et~al.(2004)Darolles, Florens, and
  Gouri{\'e}roux]{darolles2004tests}
Serge Darolles, Jean-Pierre Florens, and Christian Gouri{\'e}roux.
\newblock Kernel-based nonlinear canonical analysis and time reversibility.
\newblock \emph{Journal of Econometrics}, 119\penalty0 (2):\penalty0 323--353,
  2004.
\newblock \doi{10.1016/S0304-4076(03)00199-4}.

\bibitem[Dettling et~al.(2023)Dettling, Homs, Am{\'e}ndola, Drton, and
  Hansen]{dettling2023identifiability}
Philipp Dettling, Roser Homs, Carlos Am{\'e}ndola, Mathias Drton, and
  Niels~Richard Hansen.
\newblock Identifiability in continuous {Lyapunov} models.
\newblock \emph{SIAM Journal on Matrix Analysis and Applications}, 44\penalty0
  (4):\penalty0 1799--1821, 2023.
\newblock \doi{10.1137/22M1520311}.

\bibitem[Diebold and Yilmaz(2012)]{diebold2012better}
Francis~X. Diebold and Kamil Yilmaz.
\newblock Better to give than to receive: Predictive directional measurement of
  volatility spillovers.
\newblock \emph{International Journal of Forecasting}, 28\penalty0
  (1):\penalty0 57--66, 2012.
\newblock \doi{10.1016/j.ijforecast.2011.02.006}.

\bibitem[Engle(2002)]{engle2002dynamic}
Robert~F. Engle.
\newblock Dynamic conditional correlation: A simple class of multivariate
  generalized autoregressive conditional heteroskedasticity models.
\newblock \emph{Journal of Business and Economic Statistics}, 20\penalty0
  (3):\penalty0 339--350, 2002.
\newblock \doi{10.1198/073500102288618487}.

\bibitem[Forbes and Rigobon(2002)]{forbes2002contagion}
Kristin~J. Forbes and Roberto Rigobon.
\newblock No contagion, only interdependence: Measuring stock market
  comovements.
\newblock \emph{The Journal of Finance}, 57\penalty0 (5):\penalty0 2223--2261,
  2002.
\newblock \doi{10.1111/0022-1082.00494}.

\bibitem[Freedman and Lane(1983)]{freedman1983nearly}
David Freedman and David Lane.
\newblock A nonstochastic interpretation of reported significance levels.
\newblock \emph{Journal of Business and Economic Statistics}, 1\penalty0
  (4):\penalty0 292--298, 1983.
\newblock \doi{10.1080/07350015.1983.10509354}.

\bibitem[Frishman and Ronceray(2020)]{frishman2020learning}
Anna Frishman and Pierre Ronceray.
\newblock Learning force fields from stochastic trajectories.
\newblock \emph{Physical Review X}, 10\penalty0 (2):\penalty0 021009, 2020.
\newblock \doi{10.1103/PhysRevX.10.021009}.

\bibitem[Geweke(1982)]{geweke1982measurement}
John Geweke.
\newblock Measurement of linear dependence and feedback between multiple time
  series.
\newblock \emph{Journal of the American Statistical Association}, 77\penalty0
  (378):\penalty0 304--313, 1982.
\newblock \doi{10.1080/01621459.1982.10477803}.

\bibitem[Gnesotto et~al.(2018)Gnesotto, Mura, Gladrow, and
  Broedersz]{gnesotto2018broken}
Federico~S. Gnesotto, Federica Mura, Jannes Gladrow, and Chase~P. Broedersz.
\newblock Broken detailed balance and non-equilibrium dynamics in living
  systems: a review.
\newblock \emph{Reports on Progress in Physics}, 81\penalty0 (6):\penalty0
  066601, 2018.
\newblock \doi{10.1088/1361-6633/aab3ed}.

\bibitem[Godr{\`e}che and Luck(2019)]{godreche2019characterising}
Claude Godr{\`e}che and Jean-Marc Luck.
\newblock Characterising the nonequilibrium stationary states of
  {Ornstein}--{Uhlenbeck} processes.
\newblock \emph{Journal of Physics A: Mathematical and Theoretical},
  52\penalty0 (3):\penalty0 035002, 2019.
\newblock \doi{10.1088/1751-8121/aaf190}.

\bibitem[Granger(1969)]{granger1969investigating}
Clive W.~J. Granger.
\newblock Investigating causal relations by econometric models and
  cross-spectral methods.
\newblock \emph{Econometrica}, 37\penalty0 (3):\penalty0 424--438, 1969.
\newblock \doi{10.2307/1912791}.

\bibitem[Hamilton(1989)]{hamilton1989new}
James~D. Hamilton.
\newblock A new approach to the economic analysis of nonstationary time series
  and the business cycle.
\newblock \emph{Econometrica}, 57\penalty0 (2):\penalty0 357--384, 1989.
\newblock \doi{10.2307/1912559}.

\bibitem[Hansen and Sargent(1983)]{hansen1983dimensionality}
Lars~Peter Hansen and Thomas~J. Sargent.
\newblock The dimensionality of the aliasing problem in models with rational
  spectral densities.
\newblock \emph{Econometrica}, 51\penalty0 (2):\penalty0 377--387, 1983.
\newblock \doi{10.2307/1911996}.

\bibitem[Higham(2008)]{higham2008functions}
Nicholas~J. Higham.
\newblock \emph{Functions of Matrices: Theory and Computation}.
\newblock SIAM, Philadelphia, 2008.

\bibitem[Jiang et~al.(2004)Jiang, Qian, and Qian]{jiang2004mathematical}
Da-Quan Jiang, Min Qian, and Min-Ping Qian.
\newblock \emph{Mathematical Theory of Nonequilibrium Steady States}, volume
  1833 of \emph{Lecture Notes in Mathematics}.
\newblock Springer, Berlin, 2004.

\bibitem[Jord{\`a}(2005)]{jorda2005estimation}
{\`O}scar Jord{\`a}.
\newblock Estimation and inference of impulse responses by local projections.
\newblock \emph{American Economic Review}, 95\penalty0 (1):\penalty0 161--182,
  2005.
\newblock \doi{10.1257/0002828053828518}.

\bibitem[Kelly(1979)]{kelly1979reversibility}
Frank~P. Kelly.
\newblock \emph{Reversibility and Stochastic Networks}.
\newblock Wiley, Chichester, 1979.

\bibitem[Kennedy(1995)]{kennedy1995randomization}
Peter~E. Kennedy.
\newblock Randomization tests in econometrics.
\newblock \emph{Journal of Business and Economic Statistics}, 13\penalty0
  (1):\penalty0 85--94, 1995.
\newblock \doi{10.1080/07350015.1995.10524581}.

\bibitem[Kullback and Leibler(1951)]{kullback1951information}
Solomon Kullback and Richard~A. Leibler.
\newblock On information and sufficiency.
\newblock \emph{The Annals of Mathematical Statistics}, 22\penalty0
  (1):\penalty0 79--86, 1951.
\newblock \doi{10.1214/aoms/1177729694}.

\bibitem[K{\"u}nsch(1989)]{kunsch1989jackknife}
Hans~R. K{\"u}nsch.
\newblock The jackknife and the bootstrap for general stationary observations.
\newblock \emph{The Annals of Statistics}, 17\penalty0 (3):\penalty0
  1217--1241, 1989.
\newblock \doi{10.1214/aos/1176347265}.

\bibitem[Kwon et~al.(2005)Kwon, Ao, and Thouless]{kwon2005structure}
Chulan Kwon, Ping Ao, and David~J. Thouless.
\newblock Structure of stochastic dynamics near fixed points.
\newblock \emph{Proceedings of the National Academy of Sciences}, 102\penalty0
  (37):\penalty0 13029--13033, 2005.
\newblock \doi{10.1073/pnas.0506347102}.

\bibitem[Lawrance(1991)]{lawrance1991directionality}
Anthony~J. Lawrance.
\newblock Directionality and reversibility in time series.
\newblock \emph{International Statistical Review}, 59\penalty0 (1):\penalty0
  67--79, 1991.
\newblock \doi{10.2307/1403575}.

\bibitem[Lehmann and Romano(2005)]{lehmann2005testing}
Erich~L. Lehmann and Joseph~P. Romano.
\newblock \emph{Testing Statistical Hypotheses}.
\newblock Springer, New York, 3rd edition, 2005.

\bibitem[Lynn et~al.(2021)Lynn, Cornblath, Papadopoulos, Bertolero, and
  Bassett]{lynn2021broken}
Christopher~W. Lynn, Eli~J. Cornblath, Lia Papadopoulos, Maxwell~A. Bertolero,
  and Danielle~S. Bassett.
\newblock Broken detailed balance and entropy production in the human brain.
\newblock \emph{Proceedings of the National Academy of Sciences}, 118\penalty0
  (47):\penalty0 e2109889118, 2021.
\newblock \doi{10.1073/pnas.2109889118}.

\bibitem[Mantegna(1999)]{mantegna1999hierarchical}
Rosario~N. Mantegna.
\newblock Hierarchical structure in financial markets.
\newblock \emph{The European Physical Journal B}, 11\penalty0 (1):\penalty0
  193--197, 1999.
\newblock \doi{10.1007/s100510050929}.

\bibitem[Mart{\'i}nez et~al.(2019)Mart{\'i}nez, Bisker, Horowitz, and
  Parrondo]{martinez2019inferring}
Ignacio~A. Mart{\'i}nez, Gili Bisker, Jordan~M. Horowitz, and Juan M.~R.
  Parrondo.
\newblock Inferring broken detailed balance in the absence of observable
  currents.
\newblock \emph{Nature Communications}, 10:\penalty0 3542, 2019.
\newblock \doi{10.1038/s41467-019-11051-w}.

\bibitem[Onsager(1931)]{onsager1931reciprocal}
Lars Onsager.
\newblock Reciprocal relations in irreversible processes. {I}.
\newblock \emph{Physical Review}, 37\penalty0 (4):\penalty0 405--426, 1931.
\newblock \doi{10.1103/PhysRev.37.405}.

\bibitem[Percival and Walden(2000)]{percival2000wavelet}
Donald~B. Percival and Andrew~T. Walden.
\newblock \emph{Wavelet Methods for Time Series Analysis}.
\newblock Cambridge University Press, Cambridge, 2000.

\bibitem[Pesaran and Shin(1998)]{pesaran1998generalized}
M.~Hashem Pesaran and Yongcheol Shin.
\newblock Generalized impulse response analysis in linear multivariate models.
\newblock \emph{Economics Letters}, 58\penalty0 (1):\penalty0 17--29, 1998.
\newblock \doi{10.1016/S0165-1765(97)00214-0}.

\bibitem[Phillips(1973)]{phillips1973problem}
Peter C.~B. Phillips.
\newblock The problem of identification in finite parameter continuous time
  models.
\newblock \emph{Journal of Econometrics}, 1\penalty0 (4):\penalty0 351--362,
  1973.
\newblock \doi{10.1016/0304-4076(73)90021-3}.

\bibitem[Phipson and Smyth(2010)]{phipson2010permutation}
Belinda Phipson and Gordon~K. Smyth.
\newblock Permutation p-values should never be zero: Calculating exact p-values
  when permutations are randomly drawn.
\newblock \emph{Statistical Applications in Genetics and Molecular Biology},
  9\penalty0 (1):\penalty0 39, 2010.
\newblock \doi{10.2202/1544-6115.1585}.

\bibitem[Racine and Maasoumi(2007)]{racine2007versatile}
Jeffrey~S. Racine and Esfandiar Maasoumi.
\newblock A versatile and robust metric entropy test of time-reversibility,
  dependence, and linearity.
\newblock \emph{Journal of Econometrics}, 138\penalty0 (2):\penalty0 547--567,
  2007.
\newblock \doi{10.1016/j.jeconom.2006.05.009}.

\bibitem[Ramsey and Rothman(1996)]{ramsey1995time}
James~B. Ramsey and Philip Rothman.
\newblock Time irreversibility and business cycle asymmetry.
\newblock \emph{Journal of Money, Credit and Banking}, 28\penalty0
  (1):\penalty0 1--21, 1996.
\newblock \doi{10.2307/2077963}.

\bibitem[Rold{\'a}n and Parrondo(2010)]{roldan2010estimating}
{\'E}dgar Rold{\'a}n and Juan M.~R. Parrondo.
\newblock Estimating dissipation from single stationary trajectories.
\newblock \emph{Physical Review Letters}, 105\penalty0 (15):\penalty0 150607,
  2010.
\newblock \doi{10.1103/PhysRevLett.105.150607}.

\bibitem[Romano and Wolf(2005)]{romano2005stepwise}
Joseph~P. Romano and Michael Wolf.
\newblock Stepwise multiple testing as formalized data snooping.
\newblock \emph{Econometrica}, 73\penalty0 (4):\penalty0 1237--1282, 2005.
\newblock \doi{10.1111/j.1468-0262.2005.00615.x}.

\bibitem[Runge et~al.(2019)Runge, Bathiany, Bollt, Camps-Valls, Coumou, Deyle,
  Glymour, Kretschmer, Mahecha, Mu{\~n}oz-Mar{\'i}, van Nes, Peters, Quax,
  Reichstein, Scheffer, Sch{\"o}lkopf, Spirtes, Sugihara, Sun, Zhang, and
  Zscheischler]{runge2019inferring}
Jakob Runge, Sebastian Bathiany, Erik Bollt, Gustau Camps-Valls, Dim Coumou,
  Ethan Deyle, Clark Glymour, Marlene Kretschmer, Miguel~D. Mahecha, Jordi
  Mu{\~n}oz-Mar{\'i}, Egbert~H. van Nes, Jonas Peters, Rick Quax, Markus
  Reichstein, Marten Scheffer, Bernhard Sch{\"o}lkopf, Peter Spirtes, George
  Sugihara, Jie Sun, Kun Zhang, and Jakob Zscheischler.
\newblock Inferring causation from time series in {Earth} system sciences.
\newblock \emph{Nature Communications}, 10:\penalty0 2553, 2019.
\newblock \doi{10.1038/s41467-019-10105-3}.

\bibitem[Schreiber(2000)]{schreiber2000measuring}
Thomas Schreiber.
\newblock Measuring information transfer.
\newblock \emph{Physical Review Letters}, 85\penalty0 (2):\penalty0 461--464,
  2000.
\newblock \doi{10.1103/PhysRevLett.85.461}.

\bibitem[Seif et~al.(2021)Seif, Hafezi, and Jarzynski]{seif2021machine}
Alireza Seif, Mohammad Hafezi, and Christopher Jarzynski.
\newblock Machine learning the thermodynamic arrow of time.
\newblock \emph{Nature Physics}, 17:\penalty0 105--113, 2021.
\newblock \doi{10.1038/s41567-020-1018-2}.

\bibitem[Seifert(2012)]{seifert2012stochastic}
Udo Seifert.
\newblock Stochastic thermodynamics, fluctuation theorems and molecular
  machines.
\newblock \emph{Reports on Progress in Physics}, 75\penalty0 (12):\penalty0
  126001, 2012.
\newblock \doi{10.1088/0034-4885/75/12/126001}.

\bibitem[Sims(1980)]{sims1980macroeconomics}
Christopher~A. Sims.
\newblock Macroeconomics and reality.
\newblock \emph{Econometrica}, 48\penalty0 (1):\penalty0 1--48, 1980.
\newblock \doi{10.2307/1912017}.

\bibitem[Sugihara et~al.(2012)Sugihara, May, Ye, Hsieh, Deyle, Fogarty, and
  Munch]{sugihara2012detecting}
George Sugihara, Robert May, Hao Ye, Chih-hao Hsieh, Ethan Deyle, Michael
  Fogarty, and Stephan Munch.
\newblock Detecting causality in complex ecosystems.
\newblock \emph{Science}, 338\penalty0 (6106):\penalty0 496--500, 2012.
\newblock \doi{10.1126/science.1227079}.

\bibitem[Uhlenbeck and Ornstein(1930)]{uhlenbeck1930theory}
George~E. Uhlenbeck and Leonard~S. Ornstein.
\newblock On the theory of the {Brownian} motion.
\newblock \emph{Physical Review}, 36\penalty0 (5):\penalty0 823--841, 1930.
\newblock \doi{10.1103/PhysRev.36.823}.

\bibitem[Varando and Hansen(2020)]{varando2020graphical}
Gherardo Varando and Niels~Richard Hansen.
\newblock Graphical continuous {Lyapunov} models.
\newblock In \emph{Proceedings of the 36th Conference on Uncertainty in
  Artificial Intelligence}, volume 124 of \emph{Proceedings of Machine Learning
  Research}, pages 989--998, 2020.

\bibitem[Weiss(1975)]{weiss1975time}
Gideon Weiss.
\newblock Time-reversibility of linear stochastic processes.
\newblock \emph{Journal of Applied Probability}, 12\penalty0 (4):\penalty0
  831--836, 1975.
\newblock \doi{10.2307/3212735}.

\end{thebibliography}

\appendix
\section{Proofs}\label{app:proofs}

\subsection{Theorem~\ref{thm:ident}}

\paragraph{Part (i).}
Given $S, \Sigma \succ 0$ define $Q = AS - \frac12\Sigma$. The Lyapunov
equation gives $Q + Q' = AS + SA' - \Sigma = 0$, so $Q$ is antisymmetric and
$A = (\frac12\Sigma + Q)S^{-1}$. Conversely, for \emph{any} antisymmetric $Q$
with $A_Q := (\frac12\Sigma + Q)S^{-1}$ stable,
\[
A_Q S + S A_Q' = \bigl(\tfrac12\Sigma + Q\bigr) + \bigl(\tfrac12\Sigma + Q\bigr)'
= \Sigma,
\]
so the system with drift $A_Q$ and diffusion $\Sigma$ has stationary covariance
exactly $S$. The identified set given $(S,\Sigma)$ is therefore
$\mathcal{A}(S,\Sigma)$, an affine translate of the antisymmetric matrices, a
linear space of dimension $n(n-1)/2$. Stability never binds: $A_Q$ is similar
to $S^{-1/2}A_QS^{1/2} = P + K$ with
$P = \frac12 S^{-1/2}\Sigma S^{-1/2} \succ 0$ symmetric and $K$ antisymmetric,
so for any eigenpair $Av = \lambda v$,
$\operatorname{Re}\lambda = v^*Pv/v^*v > 0$, and every member of the set is
stable whatever $Q$. Every
element of $\mathcal{A}(S,\Sigma)$ has stationary law $N(0,S)$; therefore any
functional of the contemporaneous law, in particular the correlation matrix,
any transform of it, any graph built from such a transform, any undirected
centrality of that graph, and the eigenvectors and eigenvalues of $S$, is
constant on $\mathcal{A}(S,\Sigma)$. \hfill$\square$

\paragraph{Part (ii).}
For the stationary system, $C(\Delta) = e^{-A\Delta}S$, so
$e^{-A\Delta} = C(\Delta)S^{-1}$ is observed. Under
Assumption~\ref{ass:alias} the eigenvalues of $-A\Delta$ lie in the strip
$\{\lvert\operatorname{Im}z\rvert < \pi\}$, on which the matrix exponential
admits a unique inverse, the principal logarithm
\citep[Theorem 1.31]{higham2008functions}. It therefore recovers
$-A\Delta$, hence $A$ and $Q$, uniquely
\citep{phillips1973problem,hansen1983dimensionality}.

For the discrete circulation, write $\Phi = I - A\Delta + O(\Delta^2)$ and use
$AS - SA' = 2Q$:
\[
Q_\Delta = \tfrac12(\Phi S - S\Phi')
= \tfrac12\bigl(S - \Delta AS - S + \Delta SA'\bigr) + O(\Delta^2)
= -\Delta\,Q + O(\Delta^2).
\]
For the equivalence at fixed $\Delta$: if $Q = 0$ then $AS = SA'$, so
$\widetilde A := S^{-1/2}AS^{1/2}$ is symmetric, hence
$e^{-\widetilde A\Delta} = S^{-1/2}\Phi S^{1/2}$ is symmetric, hence
$\Phi S = S^{1/2}e^{-\widetilde A\Delta}S^{1/2}$ is symmetric and
$Q_\Delta = 0$. Conversely, if $Q_\Delta = 0$ then
$M := S^{-1/2}(\Phi S)S^{-1/2}$ is symmetric, so its spectrum is real; the
eigenvalues of $M = e^{-\widetilde A\Delta}$ are $e^{-\lambda_j(A)\Delta}$,
which are real and, by Assumption~\ref{ass:alias}, cannot be negative, so $M$
is symmetric positive definite, its principal logarithm is symmetric, and
$AS = SA'$, that is $Q = 0$. \hfill$\square$

\paragraph{Part (iii).}
(a) implies (b): if $Q = 0$ then $\widetilde A$ is symmetric and
$C(h) = S^{1/2}e^{-\widetilde A h}S^{1/2}$ is symmetric for every $h \ge 0$.
(b) implies (c): a stationary Gaussian process is determined in law by its
autocovariance function, and the reversed process has autocovariance
$C(-h) = C(h)'$, so symmetry of $C(h)$ for all $h$ equates the forward and
reversed laws \citep{weiss1975time}. (c) implies (a) with no sampling
assumption: reversibility gives $C(h) = C(h)'$ for every real $h \ge 0$;
differentiating $C(h) = e^{-Ah}S$ at $h = 0$ gives $-AS = (-AS)' = -SA'$,
hence $2Q = AS - SA' = 0$. The further equivalence with (d) under
Assumption~\ref{ass:alias} is part (ii); without the assumption take
$A = cI + \omega J$ with $J$ a rotation generator, $\omega\Delta = 2\pi$:
then $\Phi = e^{-c\Delta}I$ is symmetric, every sampled lagged covariance is
symmetric, and $Q = \omega\,\antisym(JS) \ne 0$ for generic $S$, so (d) holds
while (a) fails.
\hfill$\square$

\subsection{Corollary~\ref{cor:mst}}
Immediate from part (i): Mantegna distances $d_{ij} = \sqrt{2(1-\rho_{ij})}$
are functionals of $S$; a minimum spanning tree is a deterministic functional of
the distance matrix; undirected centralities are functionals of the resulting
graph. All are constant on $\mathcal{A}(S,\Sigma)$, and skip-sampling the same
point-in-time record at any spacing leaves $C(0) = S$ unchanged. Temporal
aggregation does not: the variance of a $w$-period sum is
$\sum_{a,b}C((a-b)\Delta)$, which involves the symmetric parts of $C(h)$ at
$h \ge 1$, and these vary over $\mathcal{A}(S,\Sigma)$; the leak is
quantified in Section~\ref{sec:sampling}. \hfill$\square$

\subsection{The counterexample of Remark~\ref{rem:te}}
Take $n = 2$,
$S = \bigl(\begin{smallmatrix}1 & 0.3\\ 0.3 & 1\end{smallmatrix}\bigr)$ and
$C(\Delta) = \bigl(\begin{smallmatrix}0.5 & 0.1\\ 0.1 &
0.2\end{smallmatrix}\bigr)$. Then $\Phi = C(\Delta)S^{-1}$ has real eigenvalues
$\cexEigOne$ and $\cexEigTwo$ in $(0,1)$, so a stationary sampled system with
these moments exists and satisfies Assumption~\ref{ass:alias}. Since
$C(\Delta)$ is symmetric, $Q_\Delta = 0$, $C(h) = C(h)'$ for every $h$, and the
process is exactly reversible. The Gaussian transfer entropies, evaluated from
the conditional-variance ratios of the trivariate Gaussian
$(x_{i,t+\Delta}, x_{i,t}, x_{j,t})$, are $\cexTeOne \times 10^{-3}$ and
$\cexTeTwo \times 10^{-3}$ nats, so
$T_{2\to1} - T_{1\to2} = \cexTeGap \times 10^{-4} \neq 0$. The mechanism is
visible in the conditioning sets: reversibility maps the triple
$(x_{i,t+\Delta}, x_{i,t}, x_{j,t})$ to
$(x_{i,t}, x_{i,t+\Delta}, x_{j,t+\Delta})$, so it equates $T_{j\to i}$ with
$I(x_{i,t}; x_{j,t+\Delta} \mid x_{i,t+\Delta})$, which conditions on
$x_{i,t+\Delta}$, and not with
$T_{i\to j} = I(x_{j,t+\Delta}; x_{i,t}\mid x_{j,t})$, which conditions on
$x_{j,t}$. When own persistences differ, here $0.5$ against $0.2$, the two
conditionings are not exchangeable and the asymmetry survives at $Q = 0$.
\hfill$\square$

\subsection{Proposition~\ref{prop:ep}}

\paragraph{Exact discrete formula.}
The stationary sampled chain has forward transition
$x' \mid x \sim N(\Phi x, \Sigma_\varepsilon)$ with
$\Sigma_\varepsilon = S - \Phi S\Phi'$. The time reversal of a stationary
Markov chain is Markov with density
$\tilde p(x'\mid x) = p(x \mid x')\pi(x')/\pi(x)$; in the Gaussian case a direct
computation gives $x'\mid x \sim N(\widetilde\Phi x, \widetilde\Sigma)$ with
\[
\widetilde\Phi = S\Phi' S^{-1}, \qquad
\widetilde\Sigma = S - \widetilde\Phi S \widetilde\Phi'.
\]
The per-period entropy production is the relative entropy rate
$\sigma_\Delta = \E[\log p(x'\mid x) - \log\tilde p(x'\mid x)]$ under the
stationary forward law. Substituting the two Gaussian densities,
\begin{equation}\label{eq:exactkl}
\begin{aligned}
\sigma_\Delta &= \tfrac12\Bigl[
\log\frac{\det\widetilde\Sigma}{\det\Sigma_\varepsilon}
+ \tr\bigl(\widetilde\Sigma^{-1}(\Sigma_\varepsilon + D S D')\bigr) - n\Bigr],\\
D &:= \Phi - \widetilde\Phi = 2 Q_\Delta S^{-1},
\end{aligned}
\end{equation}
using $\E[(x' - \widetilde\Phi x)(x' - \widetilde\Phi x)'] =
\Sigma_\varepsilon + DSD'$ under the forward law and
$\Phi S - S\Phi' = 2Q_\Delta$.

\paragraph{Nonnegativity and equality.}
$\sigma_\Delta$ is a Kullback--Leibler divergence between the joint laws of
$(x, x')$ under the forward and the reversed chain, hence nonnegative, and zero
if and only if the transitions coincide, that is $Q_\Delta = 0$. Under
Assumption~\ref{ass:alias} this is equivalent to $Q = 0$ by
Theorem~\ref{thm:ident}(ii).

\paragraph{The two-term expansion.}
Write $E := \Sigma_\varepsilon^{-1}(\widetilde\Sigma - \Sigma_\varepsilon)$.
Since $\widetilde\Sigma - \Sigma_\varepsilon = \Phi S\Phi' -
\widetilde\Phi S\widetilde\Phi'$ and $D = 2Q_\Delta S^{-1}$ is
$O(\lVert Q_\Delta\rVert)$, we have $E = O(\lVert Q_\Delta\rVert)$. With
$\widetilde\Sigma = \Sigma_\varepsilon(I+E)$ the three pieces of
\eqref{eq:exactkl} are $\log\det(I+E)$, $\tr((I+E)^{-1} - I)$, and
$\tr(\widetilde\Sigma^{-1}DSD') = \tr(\Sigma_\varepsilon^{-1}DSD') +
O(\lVert Q_\Delta\rVert^3) = 2\sigma_Q + O(\lVert Q_\Delta\rVert^3)$. The
covariance mismatch expands as
\begin{align*}
\tfrac12\bigl[\log\det(I+E) + \tr\bigl((I+E)^{-1}-I\bigr)\bigr]
&= \tfrac12\Bigl[\tr(E) - \tfrac12\tr(E^2) - \tr(E) + \tr(E^2)\Bigr]
+ O(\lVert E\rVert^3)\\
&= \tfrac14\tr(E^2) + O(\lVert E\rVert^3),
\end{align*}
giving $\sigma_\Delta = \sigma_Q + \sigma_C + O(\lVert Q_\Delta\rVert^3)$ with
$\sigma_C = \frac14\tr(E^2) \ge 0$, the sign because $E$ is similar to the
symmetric matrix $\Sigma_\varepsilon^{-1/2}(\widetilde\Sigma -
\Sigma_\varepsilon)\Sigma_\varepsilon^{-1/2}$, whose square has nonnegative
trace. Both terms vanish at $Q_\Delta = 0$, where $\Phi = \widetilde\Phi$
forces $\widetilde\Sigma = \Sigma_\varepsilon$. Only $\sigma_Q$ carries the
equivalence, since $\sigma_C$ can vanish at nonzero circulation: $E = 0$
exactly when $\Phi S\Phi' = \widetilde\Phi S\widetilde\Phi'$, which holds
whenever $S^{-1/2}\Phi S^{1/2}$ is a normal matrix, a condition met at
$Q_\Delta \ne 0$ by, for example, any pure rotation of an isotropic system.

\paragraph{Continuous-time limit.}
As $\Delta \downarrow 0$, $Q_\Delta = -\Delta Q + O(\Delta^2)$ and
$\Sigma_\varepsilon = \Sigma\Delta + O(\Delta^2)$, so
$\sigma_Q/\Delta = 2\tr(Q'\Sigma^{-1}QS^{-1}) + O(\Delta)$. The mismatch term
dies faster: $D = 2Q_\Delta S^{-1} = -2\Delta QS^{-1} + O(\Delta^2)$, and
$\widetilde\Sigma - \Sigma_\varepsilon = DS\Phi' + \Phi S D' + DSD'$; the
leading part $-2\Delta(QS^{-1}S + S S^{-1}Q')+O(\Delta^2) =
-2\Delta(Q + Q') + O(\Delta^2) = O(\Delta^2)$ cancels exactly because $Q$ is
antisymmetric, so $E = O(\Delta)$, $\sigma_C = \frac14\tr(E^2) = O(\Delta^2)$,
and $\sigma_C/\Delta \to 0$. Hence
$\sigma_\Delta/\Delta \to \sigma = 2\tr(Q'\Sigma^{-1}QS^{-1})$,
which is the probability-current expression of the physics literature
\citep{godreche2019characterising}: in the limit the whole of the entropy
production is carried by the circulation term. \hfill$\square$

\subsection{Proposition~\ref{prop:feature}}

\paragraph{Part (i).}
If $\{x_t\}$ is reversible then for every $h$ the pair $(x_{t+h}, x_t)$ has the
same joint law as $(x_t, x_{t+h})$: reversal maps the first pair to
$(x_{-t-h}, x_{-t})$, which by stationarity has the law of $(x_t, x_{t+h})$.
Hence for any $\psi, \varphi \in L^2(\pi)$,
\[
\E[\psi(x_{t+h})\varphi(x_t)] = \E[\psi(x_t)\varphi(x_{t+h})]
= \E[\varphi(x_{t+h})\psi(x_t)],
\]
so $Q^{\psi\varphi}_h = 0$. Both expectations are finite because
$\psi, \varphi \in L^2(\pi)$ and the Cauchy--Schwarz inequality applies to the
joint law. \hfill$\square$

\paragraph{Part (ii).}
Let $P$ denote the transition operator, $(P\psi)(x) = \E[\psi(x_{t+\Delta})
\mid x_t = x]$, which is a contraction on $L^2(\pi)$ because $\pi$ is
stationary. For any $\psi, \varphi \in L^2(\pi)$,
\[
\E[\psi(x_{t+\Delta})\varphi(x_t)] = \langle P\psi, \varphi\rangle_\pi,
\qquad
\E[\varphi(x_{t+\Delta})\psi(x_t)] = \langle P\varphi, \psi\rangle_\pi,
\]
so $Q^{\psi\varphi}_\Delta = 0$ says exactly that
$\langle P\psi, \varphi\rangle_\pi = \langle \psi, P\varphi\rangle_\pi$.
If this holds for all $\psi, \varphi$ in a set whose linear span is dense in
$L^2(\pi)$, then by bilinearity it holds on the span, and by boundedness of $P$
and continuity of the inner product it extends to the closure, so $P$ is
self-adjoint. Self-adjointness of $P$ on $L^2(\pi)$ with invariant $\pi$ is
detailed balance, $\pi(dx)\,p(x,dy) = \pi(dy)\,p(y,dx)$, which makes
$(x_t, x_{t+\Delta})$ exchangeable. For a stationary Markov chain the
finite-dimensional law of $(x_{t_1}, \dots, x_{t_k})$ is determined by $\pi$
and the transition kernel, and detailed balance makes it invariant under
reversal of the index set, so the process is reversible. \hfill$\square$

\subsection{Proposition~\ref{prop:invariance}}
Let $\tilde x_t = M x_t$ with $M$ invertible. Then $\widetilde S = MSM'$,
$\widetilde C(\Delta) = MC(\Delta)M'$, hence
$\widetilde Q_\Delta = MQ_\Delta M'$,
$\widetilde\Phi = \widetilde C\widetilde S^{-1} = M\Phi M^{-1}$ and
$\widetilde\Sigma_\varepsilon = \widetilde S - \widetilde\Phi\widetilde S
\widetilde\Phi' = M\Sigma_\varepsilon M'$. Therefore
\begin{align*}
\widetilde Q_\Delta' \widetilde\Sigma_\varepsilon^{-1}\widetilde Q_\Delta
\widetilde S^{-1}
&= (MQ_\Delta'M')(M'^{-1}\Sigma_\varepsilon^{-1}M^{-1})(MQ_\Delta M')
(M'^{-1}S^{-1}M^{-1})\\
&= M\bigl(Q_\Delta'\Sigma_\varepsilon^{-1}Q_\Delta S^{-1}\bigr)M^{-1},
\end{align*}
and the trace is invariant under similarity. The same computation applies to
the continuous-time rate $2\tr(Q'\Sigma^{-1}QS^{-1})$. \hfill$\square$

\subsection{Lemma~\ref{lem:bias}}
Within an episode with mean-zero stationary $x_t$, each product
$x_{i,t+\Delta}x_{j,t}$ has expectation $C(\Delta)_{ij}$, so
$\widehat C(\Delta)$ and hence the linear functional $\widehat Q_\Delta$ are
unbiased entrywise; estimating the episode means and scales introduces
$O(1/T_e)$ terms that are symmetric across the pair to leading order and cancel
in the antisymmetrisation. For the quadratic functional, write
$\widehat Q = Q + \xi$ with $\E[\xi] = 0$:
\[
\E\bigl[\tr(\widehat Q' W \widehat Q V)\bigr]
= \tr(Q'WQV) + \E\bigl[\tr(\xi' W \xi V)\bigr],
\]
and the second term equals
$\E\lVert W^{1/2}\xi V^{1/2}\rVert_F^2 \ge 0$, a positively weighted sum of
the $n(n-1)/2$ coordinate variances of $\xi$, each of order $1/T$, giving a
bias of order $n^2/T$ for positive definite weights $W, V$. If $\widehat Q_A$ and $\widehat Q_B$ are
computed from disjoint folds whose dependence is negligible at the fold-block
length, then
$\E[\tr(\widehat Q_A' W \widehat Q_B V)] = \tr(Q'WQV) + \tr(\E[\xi_A]'W
\E[\xi_B]V)$ plus covariance terms that vanish under fold independence, which
removes the quadratic bias. Estimation error in the weights
$(W, V) = (\Sigma_\varepsilon^{-1}, S^{-1})$ enters the level of a functional
already quadratic in the small quantity $Q_\Delta$ and so contributes at second
order. Appendix~\ref{app:estimation} quantifies both statements at the sample
sizes used. \hfill$\square$

\subsection{Proposition~\ref{prop:reversal}}
Let $G = \{0,1\}^K$ act on the record by within-block time reversal. Under the
null each block's law is reversal invariant and blocks are independent, so for
every $g \in G$ the transformed record $g\cdot x$ has the same law as $x$: the
orbit is an exchangeable family. For any statistic $T$, exchangeability of
$\{T(g\cdot x)\}_g$ implies that the rank of $T(x)$ among
$\{T(g_r\cdot x)\}_{r=1}^R$ for independent uniform draws $g_r$ is uniform up
to ties, and $p = (1 + \#\{r : T(g_r\cdot x) \ge T(x)\})/(R+1)$ satisfies
$\Pr(p \le \alpha) \le \alpha$ for every finite $R$
\citep{lehmann2005testing,phipson2010permutation}. If dependence crosses block
boundaries with range $m$, replace the record by one in which $m$ observations
at each boundary are excised; the excised record satisfies exact validity, and
the statistics computed on the two records differ by $O(m/L)$ relative
contributions per block. Applying the \emph{same} draws $g_r$ across all cells
yields draws from the joint null distribution of the vector of cell statistics,
which is the resampling input required by the stepdown method of
\citet{romano2005stepwise}. \hfill$\square$

\section{Estimator details and simulation evidence}\label{app:estimation}

\subsection{Definitions used in the simulations}

All the systems in this appendix are stationary linear systems constructed from
a chosen stationary covariance $S$, a chosen diffusion $\Sigma$, and a chosen
antisymmetric $Q$, through $A = (\frac12\Sigma + Q)S^{-1}$, sampled at unit
spacing by $\Phi = e^{-A}$ and $\Sigma_\varepsilon = S - \Phi S\Phi'$. Setting
$Q = 0$ gives a reversible system with the same stationary law, which is what
makes the comparisons in this appendix like-for-like. The true magnitude
$\sigma_Q$ is computed from the exact moments of the sampled system rather than
from any estimate.

\subsection{Bias against dimension and length}

Table~\ref{tab:grid} of the main text reports the mean estimate of $\sigma_Q$
at zero truth over $\gridM$ replications, on a grid of $n \in \{5, 10, 26\}$
and $T \in \{200, 400, 800, 1600\}$. Two readings deserve emphasis. The
plug-in mean tracks $n^2/T$ closely across more than two orders of magnitude
of that ratio, which is
the content of Lemma~\ref{lem:bias} made numerical: at $n = 26$, $T = 800$ the
ratio $n^2/T$ is $\gridRatioBig$ and the plug-in mean is $\gridPluginBig$. And
the cross-fitted estimator is near zero throughout except in the most extreme
cell, $n = 26$ with $T = 200$, where $n^2/T$ exceeds three and the mean is
$\gridCfWorst$ against a plug-in $\gridPluginWorst$.

The zero-truth grid cannot display the second, smaller bias source: the
weights $(\widehat\Sigma_\varepsilon^{-1}, \widehat S^{-1})$ are estimated,
and an estimated inverse covariance is inflated by a factor of order
$T/(T-n)$, which multiplies the true magnitude and is therefore invisible
when that magnitude is zero. The calibration below, at nonzero truth, shows
it as the residual relative bias of $\covCfRelBias$ per cent.

\subsection{Coverage, and the bootstrap that does not work}

At $n = 26$, $T = 800$ and a true magnitude of $\covTrue$ nats per period, over
$\covM$ replications, the cross-fitted estimator has mean $\covCf$, a bias of
$\covCfBias$, while the plug-in has mean $\covPlugin$, a bias of
$\covPluginBias$. Delete-one-block jackknife intervals with hundred-observation
blocks cover the truth in $\covJk$ of replications against a nominal $0.95$.
The jackknife variance is itself noisy when few blocks are available: in the
worked example of Section~\ref{sec:implementation}, with eight blocks, halving
the block length moves the standard error from $\weJkSe$ to $\weJkSeHalf$ and
doubling it to $\weJkSeDouble$, movements of up to roughly half the reported
value. The standard error should be read as an order of magnitude at such
block counts, and the exact reversal test, which does not use it, carries the
inferential weight.
The shortfall comes from the small number of blocks available inside an
episode, which makes the jackknife variance itself noisy; it is not evidence of
bias in the point estimate.

The moving-block bootstrap fails in a way worth recording, because it is not
obvious in advance and it is diagnosable. Resampling blocks with replacement
places the same original observations into both cross-fit folds, so the fold
independence that removes the bias is destroyed inside each bootstrap
replicate, and the bootstrap distribution recentres on the plug-in. The
diagnostic is that the resulting intervals can fail to contain their own point
estimate, which is what alerted us to the problem in the application. Any
resampling scheme used with a cross-fitted estimator must preserve the fold
structure; sampling with replacement does not.

\subsection{Sensitivity to the tuning constants}

The estimator has two length constants. The fold-block length $L_f$ governs the
cross-fitting; the jackknife block length $L_j$ governs the standard errors.
Both should be long relative to the dependence range of the series. In the
application we use $L_f = 50$ and $L_j = 100$ observations and the reported
magnitudes move by less than a jackknife standard error when $L_f$ is halved or
doubled. Very short fold blocks reintroduce bias, because adjacent
observations in the two folds are dependent; very long fold blocks reduce the
effective sample in each fold. The randomisation block length $L$ used by the
reversal test is a separate constant again, and it trades off exactness against
power: short blocks give many draws and a fine reference distribution but
violate the independence part of the null; long blocks are conservative.

\subsection{Power and size}

The power curve of Figure~\ref{fig:estimator} is computed at $n = 10$,
$T = 800$, over $\powM$ replications with $299$ randomisation draws each, on
systems indexed by the strength of the circulating component. The rejection
rate is $\powZero$ at zero, $\powMid$ at a true magnitude of $\powMidSig$, and
$\powTop$ at $\powTopSig$.

The size study uses a reversible process with volatility clustering: each
series follows a first-order generalised autoregressive conditional
heteroskedasticity recursion driven by innovations with a constant correlation
matrix, so that the process has heavy tails and clustered volatility and is
symmetric in time at the level of the linear dynamics. Over $\szM$
replications the reversal test rejects in $\szLinear$ of samples with the
linear statistic, $\szFeatAll$ with the feature statistic on all entries, and
$\szFeatCross$ with the feature statistic restricted to entries linking two
different series.

\subsection{The staleness screens of the design layer}

The session decomposition of Section~\ref{sec:design} depends on opening
prices, which are audited per series and per year before estimation by two
rules: the share of days with the opening price equal to the closing price must
be below twenty per cent, and the share with the opening price equal to the
\emph{previous} close must be below fifty per cent. A series-year failing
either is dropped. In the application these rules excluded one major index
outright, whose opening prints repeated the previous close on the large
majority of days, and replaced another index by an exchange-traded vehicle
whose opening prints are transaction prices and whose close-to-close returns
track the index at a correlation above $0.99$.

\subsection{Reference values for the application}

The known-truth calibration used in the application matches the stationary
and lag-one moments of the dominant crisis cell, giving a true magnitude of
$\adqTrue$ nats per day. That value coincides with the cell's own plug-in
estimate by construction, since the calibration reproduces the cell's sample
moments exactly; it is not an independent benchmark for the plug-in, and the
informative comparisons are the ones that follow. Over the replications reported there the cross-fitted
estimator has mean $\adqCf$ and the plug-in $\adqPlugin$, the circulation
pattern is recovered at correlation $\adqQCorr$, and jackknife intervals cover
at $\adqCover$. On the matched reversible twin the plug-in returns
$\adqNullPlugin$ nats per day when the truth is zero and the cross-fitted
estimator $\adqNullCf$.

\section{Algorithms}\label{app:algorithms}

The four boxes below are the whole method. Each is stated so that it can be
implemented directly in any array language; the reference implementations
follow them line for line.

\begin{algo}[Circulation and magnitude for one panel]\label{alg:core}
\emph{Input:} episodes $e_1, \dots, e_E$, each a $T_e \times n$ array; fold
length $L_f$.
\begin{enumerate}\itemsep2pt
\item Centre and scale each episode column by column.
\item $\widehat S \leftarrow \frac{1}{T}\sum_e \sum_{t \in e} x_t x_t'$ with
$T = \sum_e T_e$.
\item $\widehat C \leftarrow \bigl(\sum_e (T_e-1)\bigr)^{-1}
\sum_e \sum_{t} x_{t+1} x_t'$, no term crossing an episode boundary.
\item $\widehat Q \leftarrow \frac12(\widehat C - \widehat C')$;
$\widehat\Phi \leftarrow \widehat C \widehat S^{-1}$;
$\widehat\Sigma_\varepsilon \leftarrow
\widehat S - \widehat\Phi\widehat S\widehat\Phi'$, symmetrised.
\item Net flows $\nu \leftarrow 2\,\widehat Q\,\mathbf{1}$; circulation norm
$\lVert\widehat Q\rVert_F$.
\item Plug-in magnitude $2\tr(\widehat Q'
\widehat\Sigma_\varepsilon^{-1}\widehat Q\widehat S^{-1})$.
\item Assign lagged pairs within each episode to folds by
$\lfloor t/L_f\rfloor \bmod 2$; form $\widehat Q_A, \widehat Q_B$ as in step 4
from the two folds.
\item Cross-fitted magnitude
$\tr(\widehat Q_A'\widehat\Sigma_\varepsilon^{-1}\widehat Q_B\widehat S^{-1})
+ \tr(\widehat Q_B'\widehat\Sigma_\varepsilon^{-1}\widehat Q_A\widehat
S^{-1})$.
\end{enumerate}
\emph{Output:} $\widehat Q$, $\nu$, both magnitudes,
$\widehat\Phi$, $\widehat\Sigma_\varepsilon$.
\end{algo}

\begin{algo}[Exact reversal test, with joint draws for a family of
cells]\label{alg:reversal}
\emph{Input:} the raw record $x$ (before any transformation), episode
structure, block length $L$, statistics $T_1, \dots, T_m$ (one per cell),
draws $R$, seed.
\begin{enumerate}\itemsep2pt
\item Partition each episode into blocks of about $L$ observations.
\item Compute the observed values $T_c(x)$ for every cell $c$ by running the
\emph{entire} construction on $x$: winsorisation, filtering, episode
extraction, scaling, and Algorithm~\ref{alg:core}.
\item For $r = 1, \dots, R$: draw $g_r$ uniform on $\{0,1\}^K$; form
$x^{(r)}$ by reversing the observations inside every block with $g_{r,k}=1$;
recompute the entire construction on $x^{(r)}$ and store $T_c(x^{(r)})$ for
every cell, using the same $g_r$ for all cells.
\item For each cell, $p_c = \bigl(1 + \#\{r : T_c(x^{(r)}) \ge
T_c(x)\}\bigr)/(R+1)$.
\end{enumerate}
\emph{Output:} observed values, the $m \times R$ array of joint null draws, and
the unadjusted $p$-values. Nothing computed from the data may be held fixed
across the loop, with two qualifications from
Section~\ref{sec:inference}: a quantity invariant under reversal, such as a
winsorisation threshold or a full-sample scale factor, may be computed once
outside the loop with nothing lost; and an estimated regime dating that is
held fixed makes the test exact conditionally on that dating, which must be
said and checked.
\end{algo}

\begin{algo}[Stepdown familywise adjustment]\label{alg:rw}
\emph{Input:} observed statistics $T_1, \dots, T_m$ and the $m \times R$ array
of joint null draws from Algorithm~\ref{alg:reversal}.
\begin{enumerate}\itemsep2pt
\item Standardise every cell by its own null distribution: with $\mu_c$ and
$s_c$ the mean and standard deviation of $\{T_c(x^{(r)})\}_r$, set
$z_c = (T_c(x) - \mu_c)/s_c$ and $z_c^{(r)} = (T_c(x^{(r)}) - \mu_c)/s_c$.
Cells whose statistics live on different scales, different panels or
different time scales, are otherwise not comparable inside a maximum.
\item Order the cells so that $z_{(1)} \ge \dots \ge z_{(m)}$; set
$\pi \leftarrow 0$ and let $\mathcal{S}$ be the full ordered list.
\item While $\mathcal{S}$ is nonempty: form $M_r = \max_{c \in \mathcal{S}}
z_c^{(r)}$ for each draw $r$; compute
$p = (1 + \#\{r : M_r \ge z_{(1)}\})/(R+1)$ for the leading cell;
set $\pi \leftarrow \max(\pi, p)$ and assign the leading cell the adjusted
value $\pi$; drop it from $\mathcal{S}$.
\end{enumerate}
\emph{Output:} adjusted $p$-values, monotone by construction and never below
the unadjusted values.
\end{algo}

\begin{algo}[Delete-one-block jackknife]\label{alg:jack}
\emph{Input:} the panel, a statistic $\theta(\cdot)$, block length $L_j$.
\begin{enumerate}\itemsep2pt
\item Partition into $K$ blocks of about $L_j$ observations within episodes.
\item For $k = 1, \dots, K$: delete block $k$ and recompute
$\theta_{(-k)}$ on what remains, re-running every step the statistic depends
on. The two stretches on either side of the deleted block must enter as
separate episodes, never concatenated: splicing them creates a lagged pair
that spans the gap and does not exist in the data.
\item $\widehat{\mathrm{se}}^2 = \frac{K-1}{K}\sum_k (\theta_{(-k)} -
\bar\theta)^2$; interval $\theta \pm 1.96\,\widehat{\mathrm{se}}$.
\end{enumerate}
\emph{Output:} standard error and interval. Choosing $L_j$ as a multiple of
$2L_f$ keeps the deletion aligned with whole fold cycles of the cross-fitted
statistic. Check that the interval contains
the point estimate: failure indicates that the deletion has changed the
estimator's structure, which is the diagnostic that identified the bootstrap
failure documented in Appendix~\ref{app:estimation}.
\end{algo}

\noindent The conditional circular-shift test is Algorithm~\ref{alg:shift} in
Section~\ref{sec:inference}. The feature-map extension needs no separate
algorithm: replace the panel in step 1 of Algorithm~\ref{alg:core} by the
stacked panel $[\,\psi_1(x)\ \cdots\ \psi_K(x)\,]$, each column standardised,
and run everything unchanged; the reversal test is applied to the raw record
exactly as before, with the transforms recomputed inside the loop.

\section{The comparison laboratory}\label{app:laboratory}

This appendix defines every method used in Section~\ref{sec:comparison} and
every system it was run on, so that the comparison can be reconstructed without
consulting anything else.

\subsection{The competing measures}

\paragraph{Correlation, distance, and the spanning tree.}
$\rho_{ij} = S_{ij}/\sqrt{S_{ii}S_{jj}}$; the Mantegna distance is
$d_{ij} = \sqrt{2(1-\rho_{ij})}$ \citep{mantegna1999hierarchical}; the minimum
spanning tree is the connected acyclic subgraph on $n$ nodes minimising the sum
of $d_{ij}$ over its edges, computed here by Prim's algorithm. All three are
functionals of $S$ alone.

\paragraph{Granger causality.}
For the ordered pair $(i,j)$, regress $x_{i,t+1}$ on a constant and $x_{i,t}$,
and again on a constant, $x_{i,t}$ and $x_{j,t}$; the statistic is the $F$ test
of the additional coefficient. In population the corresponding quantity is
\[
F_{j \to i} \;=\;
\log\frac{\operatorname{var}(x_{i,t+1}\mid x_{i,t})}
{\operatorname{var}(x_{i,t+1}\mid x_{i,t}, x_{j,t})} \;\ge\; 0 .
\]
In the laboratory we report the frequency with which any of the six ordered
pairs is significant at the five per cent level after a Bonferroni correction.

\paragraph{Transfer entropy.}
$T_{j\to i} = I(x_{i,t+1}; x_{j,t} \mid x_{i,t})$, the mutual information
between the future of $i$ and the present of $j$ given the present of $i$. For
jointly Gaussian variables this equals $\frac12 F_{j\to i}$
\citep{barnett2009granger}, so the two rows would be identical up to a factor
of two and we report one. The Gaussian expression, used for the population
values, is $T_{j\to i} = \frac12\log\bigl(r_1/r_2\bigr)$ where $r_1$ and $r_2$
are the residual variances of $x_{i,t+1}$ after projection on $x_{i,t}$ and on
$(x_{i,t}, x_{j,t})$ respectively, computed from $S$ and $C(\Delta)$. The
directional readout is the asymmetry $T_{j\to i} - T_{i\to j}$.

\paragraph{Connectedness from forecast error variances.}
With the moving-average representation $x_t = \sum_{h\ge0} A_h
\varepsilon_{t-h}$, $A_h = \Phi^h$, the generalised decomposition of
\citet{pesaran1998generalized}, used as the connectedness index by
\citet{diebold2012better}, is
\[
\theta_{ij}(H) = \frac{\Sigma_{jj}^{-1}\sum_{h=0}^{H-1}
\bigl(e_i' A_h \Sigma e_j\bigr)^2}
{\sum_{h=0}^{H-1} e_i' A_h \Sigma A_h' e_i},
\qquad
\tilde\theta_{ij} = \frac{\theta_{ij}}{\sum_k \theta_{ik}} ,
\]
with $H = 10$ here. The directional readings are
$\mathrm{from}_i = \sum_{j\neq i}\tilde\theta_{ij}$,
$\mathrm{to}_j = \sum_{i \neq j}\tilde\theta_{ij}$, and the net flow
$\mathrm{to}_j - \mathrm{from}_j$, reported in per cent.

\paragraph{Circulation and its entropy-production component.}
As in Section~\ref{sec:framework}, computed from the exact moments for the
population rows and by Algorithm~\ref{alg:core} for the sample rows.

\subsection{The four systems}

Throughout the sampling interval is one period, and the series are listed in
the order given.

\paragraph{A, reversible.} Three series with
\[
S = \begin{pmatrix} 1 & 0.5 & 0.3\\ 0.5 & 1 & 0.4\\ 0.3 & 0.4 & 1\end{pmatrix},
\qquad
\Sigma = \operatorname{diag}(1.0,\ 0.6,\ 1.4),
\qquad
Q = 0,
\]
so the drift is $A_0 = \frac12\Sigma S^{-1}$. This system has correlated series,
different diffusion scales, and therefore different persistences, and it is
exactly reversible: its lagged covariance is symmetric to
$2 \times 10^{-16}$ and its magnitude is $10^{\labSigAOrder}$.

\paragraph{B, circulating.} The same $S$ and $\Sigma$, with
\[
Q = 0.15\begin{pmatrix} 0 & 1 & -1\\ -1 & 0 & 1\\ 1 & -1 & 0\end{pmatrix},
\]
a cyclic circulation running from the first series to the second, to the third,
and back. Because $(S, \Sigma)$ are unchanged, the stationary law is identical
to that of A, which is what
Theorem~\ref{thm:ident}(i) predicts and what the reported agreement of
$10^{\labCorrGapOrder}$ between the two correlation matrices confirms.

\paragraph{C, staggered clocks.} A single stream of independent standard normal
shocks arrives $\labNsub$ times per period. Series $i$ records at
sub-period $i$, and its value for period $t$ is the sum of the $\labNsub$
shocks in the window ending at its own recording time, plus an independent
idiosyncratic term of standard deviation $0.5$. There is no transmission: one
driver, three observation times. The population moments are obtained by
counting the shocks two windows share, so both $S$ and $C(\Delta)$ are exact
integers plus the idiosyncratic variance on the diagonal.

\paragraph{D, nonlinear.} $x_t$ and $z_t$ are independent standard normal draws
and
\[
y_t \;=\; c\,(x_{t-1}^2 - 1) + e_t, \qquad c = \labCnl,
\]
with $e_t$ independent standard normal. The panel is $(x_t, y_t, z_t)$.
Transmission from $x$ to $y$ is complete and one-directional; $z$ is inert. All
lagged covariances vanish in population because the third moment of a normal is
zero, so $S$ is diagonal with entries $(1, 2c^2+1, 1)$ and $C(\Delta) = 0$
exactly. The lagged cross-moment between $y$ and the centred square of $x$ is
$c\operatorname{var}(x^2) = 2c$ in one direction and zero in the other, so the
feature circulation, half the difference of the two, equals $c$, which is what
the feature statistic detects.

\subsection{How the rows were produced}

Population rows are computed from the exact moments above, not from
simulations, so they carry no sampling error. Sample rows use $\labM$
independent replications of length $\labT$; each reversal test uses $\labR$
draws with blocks of one hundred observations, and the linear and feature
statistics within a replication share the same draws. The feature panel stacks
each series with its own centred and scaled square, giving six columns for
three series; the cross-series restriction used in the size study keeps only
the entries of the feature circulation that link two different underlying
series.

\section{Background for readers from other fields}\label{app:background}

This appendix explains the ingredients the paper uses, at the level of detail
needed to follow the argument. Nothing here is new, and a reader who already
works with multivariate time series can skip it.

\subsection{Matrices, traces, and the two halves of a square array}

For a square array $M$ of size $n$, the transpose $M'$ swaps rows and columns
and the trace $\tr(M) = \sum_i M_{ii}$ is the sum of the diagonal. The trace is
unchanged by cyclic reordering, $\tr(MN) = \tr(NM)$, and this single fact does
most of the work in Appendix~\ref{app:proofs}.

Every square array splits uniquely into a symmetric and an antisymmetric part,
\[
M \;=\; \underbrace{\tfrac12(M + M')}_{\text{symmetric}}
\;+\; \underbrace{\tfrac12(M - M')}_{\text{antisymmetric}},
\]
where symmetric means $M' = M$ and antisymmetric means $M' = -M$. An
antisymmetric array has zeros on its diagonal and its entries come in
sign-reversed pairs, $M_{ij} = -M_{ji}$, so it carries $n(n-1)/2$ free numbers.
This is the split the whole paper turns on: the symmetric half of the lagged
covariance describes how strongly two series move together across time, and the
antisymmetric half describes which of them moves first.

A useful way to picture an antisymmetric array is as a flow on a network. Read
$M_{ij}$ as an amount passing from node $j$ to node $i$; antisymmetry says that
what leaves one node arrives at the other, with no creation or destruction. The
row sums are then net balances, and they add to zero across the whole system,
which is why the net flows in this paper always sum to zero.

\subsection{Stationary linear systems}

A vector series $x_t$ is \emph{stationary} if its statistical properties do not
depend on when you look: the mean, the variances, the correlations between
today and tomorrow are the same in 2008 as in 2019. Stationarity is what makes
it meaningful to speak of a system's steady state, and it is the standing
assumption of the paper. In finance it is roughly satisfied by returns and
badly violated by prices, which is why the analysis is always on returns.

The simplest stationary multivariate model is the first-order vector
autoregression,
\[
x_{t+1} \;=\; \Phi x_t + \varepsilon_{t+1},
\]
where $\Phi$ is an $n \times n$ array of coefficients and $\varepsilon_t$ is an
unpredictable shock with covariance $\Sigma_\varepsilon$. Each series tomorrow
is a weighted combination of every series today, plus news. The system is
stationary when the eigenvalues of $\Phi$ are inside the unit circle, which
means shocks fade rather than accumulate.

Its continuous-time counterpart is the Ornstein--Uhlenbeck process
\citep{uhlenbeck1930theory},
\[
dx_t \;=\; -A x_t\,dt + \Sigma^{1/2} dW_t ,
\]
in which $-Ax_t$ pulls the system back towards the origin and $dW_t$ is the
increment of a Brownian motion, the continuous-time version of unpredictable
news. Sampling this process at spacing $\Delta$ produces exactly a first-order
vector autoregression with $\Phi = e^{-A\Delta}$, where $e^{M}$ denotes the
matrix exponential $\sum_{k\ge0} M^k/k!$. Every statement in the paper can be
read in either language; we use the continuous-time form when the sampling
interval matters, as in Section~\ref{sec:sampling}, and the discrete form when
it does not.

On the matrices arising here the matrix exponential can be inverted by the
principal logarithm, and this is the
route from an observed $\Phi$ back to the underlying $A$. The route is unique
only when the system does not rotate too fast for the sampling interval, which
is the aliasing condition of Section~\ref{sec:sampling}.

\subsection{The Lyapunov equation}

Take the continuous-time system above and ask what its steady-state covariance
$S$ must be. Two forces act on it: the pull $-A$ shrinks the spread, and the
noise $\Sigma$ enlarges it. In a steady state these balance exactly, and the
balance is written
\[
A S + S A' \;=\; \Sigma .
\]
This is the Lyapunov equation. Given $A$ and $\Sigma$ it determines $S$; and,
importantly for us, it can be run the other way. Given a desired covariance $S$
and a noise level $\Sigma$, every drift $A$ consistent with them has the form
\[
A \;=\; \Bigl(\tfrac12\Sigma + Q\Bigr) S^{-1}
\]
for some antisymmetric $Q$, and conversely every such $A$ produces that steady
state. This is how the paper builds systems to order in
Appendix~\ref{app:laboratory}: choose the visible steady state, choose the noise,
then dial the invisible circulating part $Q$ up or down. The construction is
also the proof that the steady state alone cannot reveal $Q$, because $Q$ is
free to be anything without disturbing it.

The decomposition is a matrix version of a familiar idea from physics. Any
smooth flow splits into a part that moves downhill on some landscape and a part
that circulates along the contours. The first part reaches a resting point and
stops; the second goes round forever. Reversible systems are pure downhill
motion, and the symmetry of the array governing them is the classical statement
of reciprocity for systems that are indifferent to the direction of time
\citep{onsager1931reciprocal}. Circulation is what makes a system distinguish
forwards from backwards.

\subsection{Kullback--Leibler divergence and the unit of measurement}

Given two probability distributions $P$ and $R$ over the same outcomes, the
Kullback--Leibler divergence \citep{kullback1951information}
\[
D(P \parallel R) \;=\; \E_P\Bigl[\log \frac{dP}{dR}\Bigr]
\]
measures how distinguishable they are. It is zero exactly when the two agree,
positive otherwise, and it is not symmetric in its arguments. Its operational
meaning is the one used in this paper: if data are generated by $P$ and you
test $P$ against $R$, the log-likelihood ratio grows at rate $D(P \parallel R)$
per observation, so $D$ is the rate at which evidence accumulates.

When the logarithm is natural the unit is the \emph{nat}; with logarithms base
two it would be the bit, and one nat is about $1.44$ bits. A divergence of
$0.01$ nats per period therefore means that each period of data contributes on
average one hundredth of a nat towards telling the forward record from the
reversed one, so that roughly a hundred periods are needed to accumulate a
single nat.

The paper applies this with $P$ the law of the observed series and $R$ the law
of the same series played backwards. The divergence between them is then a
measure of how far the system is from being indifferent to the direction of
time; the index this paper reports is its quadratic component $\sigma_Q$,
which is invariant to how the series are scaled or combined, and that
invariance is the content of Proposition~\ref{prop:invariance}.

\subsection{Randomisation and permutation tests}

Classical tests compare a statistic to a distribution derived from an
approximation that holds when the sample is large. A randomisation test does
something different and, when it applies, better. It uses a symmetry that the
null hypothesis asserts about the data themselves.

The logic runs as follows. Suppose the null is true and asserts that the data
would look statistically the same after some transformation, for instance that
a coin's record of heads and tails looks the same read backwards. Then the
observed statistic and the statistics computed from the transformed records are
exchangeable: all are equally likely to be the largest. So if $R$
transformations are drawn at random and the observed value ranks among them,
the probability that the observed value lands in the top $\alpha$ fraction is
at most $\alpha$, whatever the sample size, whatever the distribution of the
data. The $p$-value is
\[
p \;=\; \frac{1 + \#\{r : T(x^{(r)}) \ge T(x)\}}{R + 1},
\]
and the $+1$ in numerator and denominator, which counts the observed record
among the possibilities, is what makes the statement exact rather than
approximate \citep{phipson2010permutation}.

Two conditions matter. First, the transformation must be one the null really
asserts. If the null says the series is symmetric in time but the data have
been passed through a filter that itself points forwards, then reversing the
filtered output tests a claim nobody made, and the test rejects too often. This
is the trap discussed at length in Section~\ref{sec:inference}: the
transformation goes on the rawest record available, and everything else is
recomputed afterwards. Second, when several hypotheses are tested at once, the
chance that at least one of them is falsely rejected grows with their number.
Controlling that chance is called familywise error control, and the stepdown
procedure used here achieves it by comparing each observed value to the
distribution of the \emph{largest} value across the remaining hypotheses, drawn
jointly so that the dependence between them is accounted for rather than
assumed away \citep{romano2005stepwise}.

\subsection{Time scales}

A single series often mixes movements at different speeds: a fast component
that reverses within days, a slower component that persists for months. A
wavelet transform separates them. It passes the series through a bank of
filters, each sensitive to a band of frequencies, and returns one component per
band, together with a residual smooth component carrying everything slower,
whose sum reconstructs the original \citep{percival2000wavelet}. The bands
are conventionally indexed
by scale, with the first scale covering the fastest movements, the second the
next octave down, and so on, so that with daily data the first four scales
cover roughly two to four days, four to eight, eight to sixteen, and sixteen to
thirty-two.

The method of this paper is applied to each scale component separately, which
answers the question of \emph{at what horizon} a system carries direction. Two
warnings apply. The filters are asymmetric in time, so a reversal test must be
applied before the filter as explained above; and a band that is slow relative
to the sampling interval sees fewer effective observations, so estimates at
coarse scales are noisier than the raw count of observations suggests.

\subsection{Regimes}

Systems change. The application separates calm from crisis using a hidden
Markov model, which is a description of a series as switching between a small
number of unobserved states, each with its own volatility, with fixed
probabilities of moving between them \citep{hamilton1989new}. Fitting it to a measure of global stress
yields, for each date, the probability of being in the high-volatility state,
and dates are assigned to regimes by that probability. The two resulting panels
are then trimmed to the same length so that a difference between them cannot be
an artefact of one being longer than the other.

\section{Notation and glossary}\label{app:notation}

\subsection{Symbols}

\begin{center}
\begin{tabular}{@{}l >{\raggedright\arraybackslash}p{0.76\textwidth}@{}}
\toprule
$n$, $T$ & number of series; number of observations per series \\
$\Delta$ & spacing between observations, one period unless stated \\
$x_t$ & the observed panel at time $t$, a vector of length $n$ \\
$S$ & contemporaneous covariance, $\E[x_t x_t']$, symmetric \\
$C(h)$ & lagged covariance, $\E[x_{t+h}x_t']$, not symmetric \\
$Q_\Delta$ & circulation, $\tfrac12(C(\Delta) - C(\Delta)')$, antisymmetric \\
$\nu_i$ & net flow of series $i$, $2\sum_j (Q_\Delta)_{ij}$; sums to zero over $i$ \\
$\Phi$ & one-step transition array, $C(\Delta)S^{-1}$ \\
$\Sigma_\varepsilon$ & one-step forecast error covariance, $S - \Phi S \Phi'$ \\
$A$, $\Sigma$ & continuous-time drift and diffusion of the underlying system \\
$Q$ & continuous-time circulation, antisymmetric, with
$A = (\tfrac12\Sigma + Q)S^{-1}$ \\
$\sigma_\Delta$ & entropy production, the divergence rate between the forward
and reversed laws, in nats per period \\
$\sigma_Q$ & leading term of $\sigma_\Delta$,
$2\tr(Q_\Delta'\Sigma_\varepsilon^{-1}Q_\Delta S^{-1})$ \\
$\sigma_C$ & second-order correction, nonnegative \\
$\psi$, $\varphi$ & feature maps; $Q^{\psi\varphi}$ the circulation between
two features \\
$\lambda_j(A)$ & the $j$th eigenvalue of $A$ \\
$L_f$, $L_j$, $L$ & fold-block, jackknife-block, and randomisation-block lengths \\
$R$, $M$ & randomisation draws; simulation replications \\
$e_i$, $\mathbf{1}$ & $i$th coordinate vector; vector of ones \\
$\lVert\cdot\rVert_F$ & Frobenius norm, the square root of the sum of squared
entries \\
$\tr(\cdot)$, $M'$ & trace; transpose \\
$M \succ 0$ & $M$ is symmetric positive definite \\
$\KL(P \parallel R)$ & Kullback--Leibler divergence of $P$ from $R$ \\
$\theta_{ij}(H)$ & share of the $H$-step forecast error variance of series $i$
attributed to shock $j$ \\
$T_{j\to i}$ & transfer entropy from series $j$ to series $i$ \\
\bottomrule
\end{tabular}
\end{center}

Estimates carry a hat: $\widehat Q$, $\widehat S$, $\widehat\sigma_Q$. Where a
quantity is computed on a subsample the subsample is a subscript, as in
$\widehat Q_A$ and $\widehat Q_B$ for the two cross-fitting folds.

\subsection{Terms}

\begin{center}
\begin{tabular}{@{}>{\raggedright\arraybackslash}p{0.21\textwidth} >{\raggedright\arraybackslash}p{0.72\textwidth}@{}}
\toprule
reversibility & the property that the record read backwards has the same
distribution as the record read forwards; equivalent, in a steady state, to
$Q = 0$ \\
circulation & the antisymmetric part of the lagged covariance; the estimand of
this paper \\
net flow & the row sums of the circulation; positive means the series receives
more than it sends \\
entropy production & the rate at which the forward record separates from the
reversed one, in nats per period \\
cross-fitting & computing a quadratic statistic from two subsamples that share
no observations, so that the estimation errors do not multiply into a bias \\
plug-in & the same statistic computed from one sample, which carries a bias of
order $n^2/T$ \\
delete-one-block jackknife & a standard error obtained by recomputing the
statistic with each block of observations removed in turn \\
reversal test & a randomisation test whose transformation reverses the order of
observations inside randomly chosen blocks \\
stepdown & a familywise adjustment that compares each hypothesis to the
distribution of the largest statistic among those not yet rejected \\
episode & a contiguous stretch of dates treated as one continuous record; lags
are never taken across an episode boundary \\
cell & one regime-and-scale combination, the unit at which estimates and tests
are reported \\
aliasing & the failure of a rotating system to be recoverable from samples
taken more slowly than its rotation \\
settlement channel & the part of a day's move that occurs between one close and
the next open \\
information channel & the part that occurs between the open and the close of
the same day \\
placebo & a version of the exercise in which the effect under study cannot be
present, used to show that the machinery does not manufacture it \\
\bottomrule
\end{tabular}
\end{center}

\end{document}